\documentclass[twocolumn]{aastex631}

\usepackage{graphicx}
\usepackage{amsmath}
\usepackage{verbatim}

\begin{document}

\title{{exoALMA XII} : Weighing and sizing exoALMA disks with rotation curve modelling}

\author[0000-0002-0786-7307]{Cristiano Longarini}
\affiliation{Institute of Astronomy, University of Cambridge, Madingley Road, Cambridge, CB3 0HA, United Kingdom}
\affiliation{Dipartimento di Fisica, Università degli Studi di Milano, Via Celoria 16, Milano, 20133, Italy}
\correspondingauthor{Cristiano Longarini}
\email{cl2000@cam.ac.uk}

\author[0000-0002-2357-7692]{Giuseppe Lodato}
\affiliation{Dipartimento di Fisica, Università degli Studi di Milano, Via Celoria 16, Milano, 20133, Italy}

\author[0000-0003-4853-5736]{Giovanni Rosotti}
\affiliation{Dipartimento di Fisica, Università degli Studi di Milano, Via Celoria 16, Milano, 20133, Italy}

\author[0000-0003-2253-2270]{Sean Andrews}
\affiliation{Center for Astrophysics | Harvard \& Smithsonian, Cambridge, MA 02138, USA}

\author[0000-0002-7501-9801]{Andrew Winter}
\affiliation{Universit\'{e} C\^{o}te d'Azur, Observatoire de la C\^{o}te d'Azur, CNRS, Laboratoire Lagrange, 06300 Nice, France}
\affiliation{Max-Planck Institute for Astronomy (MPIA), Königstuhl 17, 69117 Heidelberg, Germany}

\author[0000-0002-0491-143X]{Jochen Stadler}
\affiliation{Universit\'{e} C\^{o}te d'Azur, Observatoire de la C\^{o}te d'Azur, CNRS, Laboratoire Lagrange, 06300 Nice, France}
\affiliation{Université Grenoble Alpes, CNRS, IPAG, 38000 Grenoble, France}

\author[0000-0001-8446-3026]{Andrés Izquierdo}
\altaffiliation{NASA Hubble Fellowship Program Sagan Fellow}
\affiliation{Department of Astronomy, University of Florida, Gainesville, FL 32611, USA}
\affiliation{Leiden Observatory, Leiden University, P.O. Box 9513, NL-2300 RA Leiden, The Netherlands}
\affiliation{European Southern Observatory, Karl-Schwarzschild-Str. 2, D-85748 Garching bei Munchen, Germany}

\author[0000-0002-5503-5476]{Maria Galloway-Sprietsma}
\affiliation{Department of Astronomy, University of Florida, Gainesville, FL 32611, USA}

\author[0000-0003-4689-2684]{Stefano Facchini}
\affiliation{Dipartimento di Fisica, Università degli Studi di Milano, Via Celoria 16, Milano, 20133, Italy}

\author[0000-0003-2045-2154]{Pietro Curone}
\affiliation{Dipartimento di Fisica, Università degli Studi di Milano, Via Celoria 16, Milano, 20133, Italy}
\affiliation{Departamento de Astronomía, Universidad de Chile, Camino El Observatorio 1515, Las Condes, Santiago, Chile}

\author[0000-0002-7695-7605]{Myriam Benisty}
\affiliation{Universit\'{e} C\^{o}te d'Azur, Observatoire de la C\^{o}te d'Azur, CNRS, Laboratoire Lagrange, 06300 Nice, France}
\affiliation{Max-Planck Institute for Astronomy (MPIA), Königstuhl 17, 69117 Heidelberg, Germany}

\author[0000-0003-1534-5186]{Richard Teague}
\affiliation{Department of Earth, Atmospheric, and Planetary Sciences, Massachusetts Institute of Technology, Cambridge, MA 02139, USA}

\author[0000-0001-7258-770X]{Jaehan Bae}
\affiliation{Department of Astronomy, University of Florida, Gainesville, FL 32611, USA}

\author[0000-0001-6378-7873]{Marcelo Barraza-Alfaro}
\affiliation{Department of Earth, Atmospheric, and Planetary Sciences, Massachusetts Institute of Technology, Cambridge, MA 02139, USA}

\author[0000-0002-2700-9676]{Gianni Cataldi}
\affiliation{National Astronomical Observatory of Japan, Osawa 2-21-1, Mitaka, Tokyo 181-8588, Japan}

\author[0000-0002-1483-8811]{Ian Czekala}
\affiliation{School of Physics \& Astronomy, University of St. Andrews, North Haugh, St. Andrews KY16 9SS, UK}

\author[0000-0003-3713-8073]{Nicolás Cuello}
\affiliation{Université Grenoble Alpes, CNRS, IPAG, 38000 Grenoble, France}

\author[0000-0003-4679-4072]{Daniele Fasano}
\affiliation{Universit\'{e} C\^{o}te d'Azur, Observatoire de la C\^{o}te d'Azur, CNRS, Laboratoire Lagrange, 06300 Nice, France}
\affiliation{Max-Planck Institute for Astronomy (MPIA), Königstuhl 17, 69117 Heidelberg, Germany}

\author[0000-0002-9298-3029]{Mario Flock} 
\affiliation{Max-Planck Institute for Astronomy (MPIA), Königstuhl 17, 69117 Heidelberg, Germany}

\author[0000-0003-1117-9213]{Misato Fukagawa}
\affiliation{National Astronomical Observatory of Japan, 2-21-1 Osawa, Mitaka, Tokyo 181-8588, Japan}

\author[0000-0002-5910-4598]{Himanshi Garg}
\affiliation{School of Physics and Astronomy, Monash University, VIC 3800, Australia}

\author[0000-0002-8138-0425]{Cassandra Hall}
\affiliation{Department of Physics and Astronomy, The University of Georgia, Athens, GA 30602, USA}
\affiliation{Center for Simulational Physics, The University of Georgia, Athens, GA 30602, USA}
\affiliation{Institute for Artificial Intelligence, The University of Georgia, Athens, GA 30602, USA}

\author[0000-0003-1502-4315]{Iain Hammond}
\affiliation{School of Physics and Astronomy, Monash University, VIC 3800, Australia}

\author[0009-0003-7403-9207]{Caitlyn Hardiman}
\affiliation{School of Physics and Astronomy, Monash University, VIC 3800, Australia}

\author[0000-0001-7641-5235]{Thomas Hilder}
\affiliation{School of Physics and Astronomy, Monash University, VIC 3800, Australia}

\author[0000-0001-6947-6072]{Jane Huang}
\affiliation{Department of Astronomy, Columbia University, 538 W. 120th Street, Pupin Hall, New York, NY 10027, United States of America}

\author[0000-0003-1008-1142]{John D. Ilee}
\affiliation{School of Physics and Astronomy, University of Leeds, Leeds, UK, LS2 9JT}

\author[0000-0001-8061-2207]{Andrea Isella}
\affiliation{Department of Physics and Astronomy, Rice University, 6100 Main St, Houston, TX 77005, USA}
\affiliation{Rice Space Institute, Rice University, 6100 Main St, Houston, TX 77005, USA}

\author[0000-0001-7235-2417]{Kazuhiro Kanagawa} 
\affiliation{College of Science, Ibaraki University, 2-1-1 Bunkyo, Mito, Ibaraki 310-8512, Japan}

\author[0000-0002-8896-9435]{Geoffroy Lesur} 
\affiliation{Univ. Grenoble Alpes, CNRS, IPAG, 38000 Grenoble, France}

\author[0000-0002-8932-1219]{Ryan A. Loomis}
\affiliation{National Radio Astronomy Observatory, 520 Edgemont Rd., Charlottesville, VA 22903, USA}

\author[0000-0002-1637-7393]{Francois Ménard}
\affiliation{Université Grenoble Alpes, CNRS, IPAG, 38000 Grenoble, France}

\author[0000-0003-4039-8933]{Ryuta Orihara}
\affiliation{College of Science, Ibaraki University, 2-1-1 Bunkyo, Mito, Ibaraki 310-8512, Japan}

\author[0000-0001-5907-5179]{Christophe Pinte}
\affiliation{Université Grenoble Alpes, CNRS, IPAG, 38000 Grenoble, France}
\affiliation{School of Physics and Astronomy, Monash University, VIC 3800, Australia}

\author[0000-0002-4716-4235]{Daniel Price}
\affiliation{School of Physics and Astronomy, Monash University, VIC 3800, Australia}

\author[0000-0003-1859-3070]{Leonardo Testi}
\affiliation{Dipartimento di Fisica e Astronomia, Universita' di Bologna, I-40190 Bologna, Italy}

\author[0000-0002-3468-9577]{Gaylor Wafflard-Fernandez}
\affiliation{Univ. Grenoble Alpes, CNRS, IPAG, 38000 Grenoble, France}

\author[0000-0002-7212-2416]{Lisa Wölfer}
\affiliation{Department of Earth, Atmospheric, and Planetary Sciences, Massachusetts Institute of Technology, Cambridge, MA 02139, USA}

\author[0000-0003-1412-893X]{Hsi-Wei Yen}
\affiliation{Academia Sinica Institute of Astronomy \& Astrophysics, 11F of Astronomy-Mathematics Building, AS/NTU, No.1, Sec. 4, Roosevelt Rd, Taipei 106216, Taiwan}

\author[0000-0001-8002-8473]{Tomohiro C. Yoshida}
\affiliation{National Astronomical Observatory of Japan, 2-21-1 Osawa, Mitaka, Tokyo 181-8588, Japan}
\affiliation{Department of Astronomical Science, The Graduate University for Advanced Studies, SOKENDAI, 2-21-1 Osawa, Mitaka, Tokyo 181-8588, Japan}

\author[0000-0001-9319-1296]{Brianna Zawadzki}
\affiliation{Department of Astronomy, Van Vleck Observatory, Wesleyan University, 96 Foss Hill Drive, Middletown, CT 06459, USA}



\begin{abstract}

The exoALMA large program offers a unique opportunity to investigate the fundamental properties of protoplanetary disks, such as their masses and sizes, providing important insights in the mechanism responsible for the transport of angular momentum. In this work, we model the rotation curves of CO isotopologues $^{12}$CO and $^{13}$CO of ten sources within the exoALMA sample, and we constrain the stellar mass, the disk mass and the density scale radius %(i.e. the radius of the exponential fall of the surface density) 
through precise characterization of the pressure gradient and disk self gravity. 
%These quantities are crucial in understanding protoplanetary disk evolution and planet formation. 
%Precise modeling of the rotation curves provides a reliable estimate of the stellar mass, considering the effects of the pressure gradient and disk self-gravity. 
We obtain dynamical disk masses for our sample measuring the self-gravitating contribution to the gravitational potential. We are able to parametrically describe their surface density, and all of them appear gravitationally stable. % Furthermore, rotation curves allow us to weigh protoplanetary disks, measuring the self-gravitating contribution to the gravitational potential. 
By combining dynamical disk masses with dust continuum emission data, we determine an averaged gas-to-dust ratio of approximately 400, not statistically consistent with the standard value of 100, assuming optically thin dust emission. In addition, the measurement of the dynamical scale radius allows for direct comparison with flux-based radii of gas and dust. This comparison suggests that substructures may influence the size of the dust disk, and that CO depletion might reconcile our measurements with thermochemical models. Finally, with the stellar mass, disk mass, scale radius, and accretion rate, and assuming self-similar evolution of the surface density, we constrain the effective $\alpha_S$ for these systems. We find a  broad range of $\alpha_S$ values ranging between $10^{-5}$ and $10^{-2}$.

\end{abstract}



\section{Introduction} \label{intro}

%Our understanding of the physical properties of protoplanetary disks has been revolutionized during the last few years thanks to the Atacama Large Millimeter Array (ALMA) \citep{alma_hltau15,andrews18,oberg21}. Indeed, high spatial and spectral resolution observations of such environments enable investigation of their temperature, velocity and density, building a comprehensive understanding of their structure \citep{ddcollab20}.  Nowadays, kinematic characterization of molecular line emission from planet-forming environments is possible \citep{pinte24}, providing an alternative to dust emission for observing planet-forming disks.

A particularly important quantity in protoplanetary disc kinematics is the rotation curve, i.e. the azimuthally averaged rotational velocity as a function of radius. Indeed, in a protoplanetary disk, the dominant motion is azimuthal \citep{pinte24}, hence having a thorough model of the rotation curve is crucial to precisely characterize such environments. On top of the standard Keplerian rotation, there are additional effects that globally influence the rotation curve, such as the pressure gradient and the disk self gravity, whose strength are connected to  fundamental disk quantities. Recently, significant work has been done to characterize rotation curves in protoplanetary disks. Leveraging the analytical work by \cite{bertin99}, \cite{veronesi21} constrained the star and disk masses of the protoplanetary disk Elias 2-27 from the rotation curves of $^{13}$CO and C$^{18}$O, marking the first dynamical estimate of a protoplanetary disk mass. Subsequent developments have introduced new methods for extracting rotation curves \citep[e.g.,][]{discminer1,discminer2}, enhancing the quality of modeling. \cite{Lodato23} presented a model for the rotation curve of a vertically isothermal disk including self-gravity, applying it to IM Lup and GM Aur. Following this, \cite{martire24} generalized the model for a vertically stratified disk and applied it to the MAPS sample. Most recently, \cite{veronesi24,Andrews24} studied the uncertainties related to this method, finding that the precision of disk mass measurements is around 25\%, with the minimum measurable mass being 5\% the stellar mass, and that it is possible to constrain surface density profile, rather than just the integrated mass.

In this work, we model the rotation curves of the sample described in \cite{Stadler_exoALMA}, extracted with \textsc{discminer} \citep{Izquierdo_exoALMA}, to constrain stellar masses, disk masses and scale radii, following the approach of the aforementioned papers. This paper is organized as follows. In section \ref{S2} we briefly present the physical model we adopt to describe the rotation curve. We use a thermally stratified model \citep{martire24}, where the thermal structure is obtained in \cite{Galloway_exoALMA}. In section \ref{S3} we present the analysis procedure, we justify the sample we are analyzing and we discuss how we treat the systematic uncertainties. In section \ref{S4} we present the results and we separately discuss disk masses, stellar masses, scale radii and the properties we can extract from them. Finally, in section \ref{conclusions} we summarize the findings and draw the conclusions.

\section{Physical model} \label{S2}
The model we adopt to describe the rotation curve of a protoplanetary disk is the one presented in \cite{Lodato23}, then generalized by \cite{martire24} including the disk thermal stratification. The need for including thermal stratification is justified in \cite{Stadler_exoALMA}, showing that the shift between $^{12}$CO and $^{13}$CO rotation curves can not be explained by a vertical isothermal model. The full description of the stratified model is given in Appendix \ref{appenidx_model}.

\subsection{2D temperature structure}
There is observational evidence that protoplanetary disks have a vertical temperature gradient \citep{Dartois03,Rosenfeld2013,Pinte18}. The thermal structure can be probed through optically thick molecular line emission \citep{MAPSIV}. \cite{Galloway_exoALMA} obtained the 2D thermal structures of exoALMA disks from $^{12}$CO and $^{13}$CO datacubes, adopting the Dartois \citep{Dartois03} prescription
\begin{equation}\label{dartois_eqt}
    T(R,z) = \left\{ \begin{array}{ll}
         T_\text{atm} + \left(T_\text{mid} - T_\text{atm} \right) \cos^2 \left(\frac{\pi z}{2Z_q}  \right), & z<Z_q \\
         T_\text{atm}, & z>Z_q
    \end{array}
    \right. 
\end{equation}
where
\begin{equation}
    \begin{array}{l}
        T_\text{atm}(R)=  T_\text{atm,100}(R/100\text{au})^{q_\text{atm}} \\
        T_\text{mid}(R)=  T_\text{mid,100}(R/100\text{au})^{q_\text{mid}} \\
        Z_q = Z_0 (R/100\text{au})^{\beta} .
    \end{array}
\end{equation}
Hence, the 2D thermal structure is described by 6 parameters, namely $T_\text{mid,100},T_\text{atm,100},q_\text{mid}, q_\text{atm}, Z_0, \beta$. In this work, we will use the Dartois prescription with the best fit values for the thermal parameters obtained by \cite{Galloway_exoALMA}, that are summarized in Appendix \ref{appenidx_paramters}.

\subsection{Model for the rotation curve}
In this paragraph we present the fundamental equations of the stratified model from \cite{martire24}. The complete derivation of the rotation curve is given in Appendix \ref{appenidx_model}.

Under the hypothesis of self-similar surface density profile \citep{Lyndenbell74}
\begin{equation}
    \Sigma(R) = \frac{(2-\gamma)M_d}{2\pi R_c^2} \left(\frac{R}{R_c}\right)^{-\gamma}\exp\left[-\left(\frac{R}{R_c}\right)^{2-\gamma}\right],
\end{equation}
and assuming a temperature structure
\begin{equation}
    T(R,z) = T_\text{mid}(R)f(R,z),
\end{equation}
with $f$ given by Eq. \eqref{dartois_eqt}. \cite{martire24} showed that the rotation curve is given by 
\begin{equation}\label{rotationcurve_strat_eq}\begin{split}
    v_\phi^2 = v_\text{k}^2 \left\{\left[1+\left(\frac{z}{R}\right)^2\right]^{-3/2} - \left[\gamma^\prime + (2-\gamma)\left(\frac{R}{R_c}\right)^{2-\gamma}  - \right. \right. \\ \left.\left. - \frac{\text{d}\log(fg)}{\text{d}\log R}\right]\left(\frac{H}{R}\right)_\text{mid}^2 f(R,z)   \right\} + v_d^2,
\end{split}\end{equation}
where  $\gamma^\prime = \gamma + (3+q_\text{mid})/2$, $v_\text{k} = \sqrt{GM_\star/R}$,
\begin{equation}
    \log(fg) = -\frac{1}{H_\text{mid}^2}\int_0^z \frac{z'}{f}\left[1+\left(\frac{z'}{R}\right)^2\right]^{-3/2} \text{d}z'.
\end{equation}
and 
\begin{equation}
\begin{gathered}
	v_d^2 = G \int^\infty_0 \Bigg[K(k) - \frac{1}{4}\Bigg(\frac{k^2}{1-k^2}\Bigg)\times \\ \times
    \Bigg(\frac{R^\prime}{R}-\frac{R}{r}+\frac{z^2}{RR^\prime}\Bigg) E(k)\Bigg]\sqrt{\frac{R^\prime}{R}} k\Sigma(R^\prime) dR^\prime,
\end{gathered}
\end{equation}
where $K(k)$ and $E(k)$ are complete elliptic integrals \citep{abramowitz} and $k^2 =  4RR^\prime/[(R + R^\prime)^2 + z^2]$.% Equation \eqref{rotationcurve_strat_eq} will be used to fit for $^{12}$CO and $^{13}$CO rotation curves simultaneously, to constrain disc mass, stellar mass and scale radius.

{The model presented in this section does not account for pressure-modulated substructures that may be present in the data. To assess their influence on the best-fit parameters, we tested a rotation curve model that includes the contribution of substructures in the pressure gradient. Specifically, we modelled the density variations associated with gaps and rings as Gaussians and then computed the pressure gradient and disc self-gravity  self-consistently. The corresponding velocity perturbations are of the order of 50-70 m/s. 
We applied this approach to the rotation curve of a reference disc, LkCa 15, fitting the data both with and without the inclusion of substructures. Our results show that the most affected parameter is the scale radius, as it is directly influenced by the pressure gradient. However, the correction remains within the typical uncertainties. In contrast, the impact of substructures on the dynamical mass estimates is minimal.}

\section{Analysis}\label{S3}

\subsection{Sample}
To apply this method, the crucial criteria are the availability of well-defined and precisely measured CO emitting surfaces along which to appropriately sample the rotation curve, and the absence of non-axisymmetric features, which may affect the azimuthal averaging, intrinsic to the definition of rotation curve, and the very assumption of centrifugal balance. This is important to ensure a correct extraction of the rotation curve, of the thermal structures and to evaluate at the right position $(R,z)$ the model of Eq. \eqref{rotationcurve_strat_eq}. Hence, it is not feasible to apply this method to the entire exoALMA sample. We discarded sources that show strong non-axisymmetric features (MWC 758, CQ Tau) and low inclination ones (HD 135344B, HD 143006, J1604), for which the extraction of the emitting layer is not well defined. Therefore, the sample used in this work includes AA Tau, DM Tau, HD 34282, J1615, J1842, J1852, LkCa15, PDS66, SY Cha and V4046 Sgr. 

\begin{figure*}
    \centering
   \includegraphics[width=\textwidth]{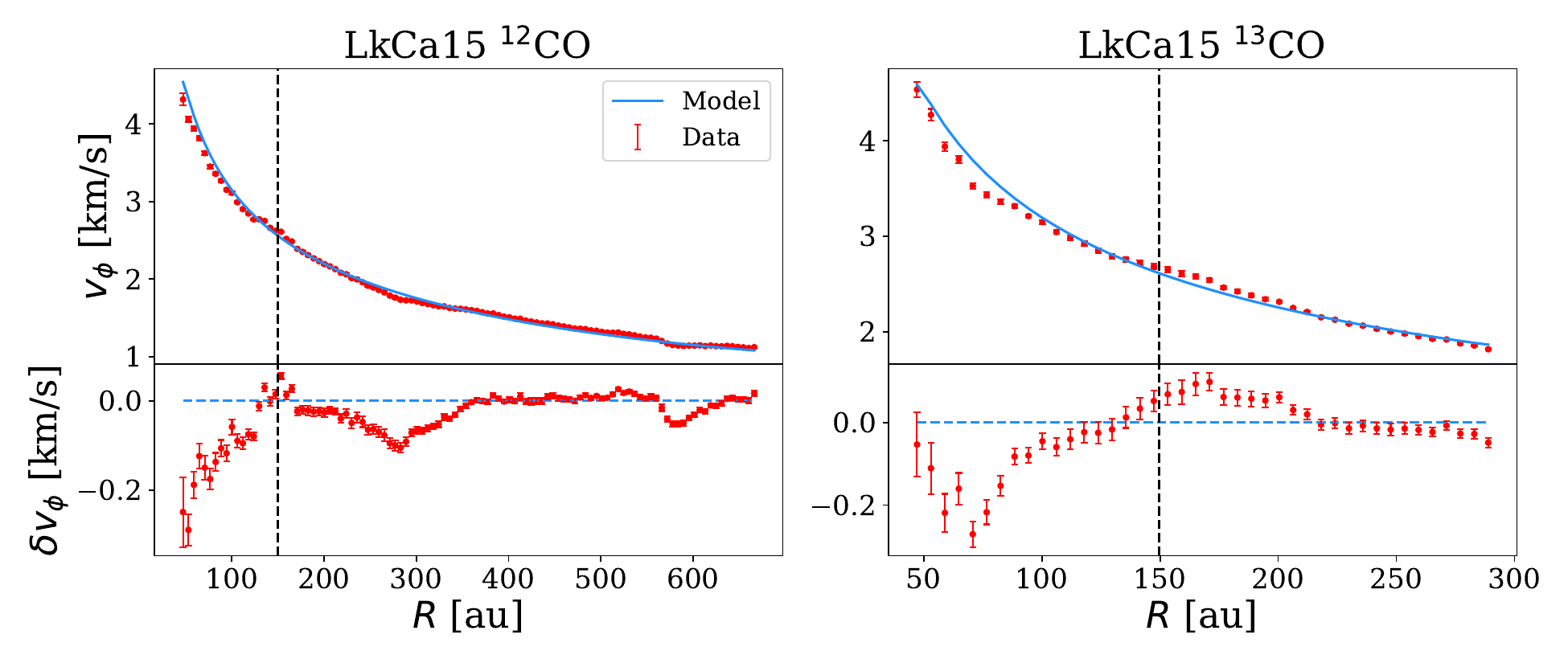}
   \includegraphics[width=0.5\textwidth]{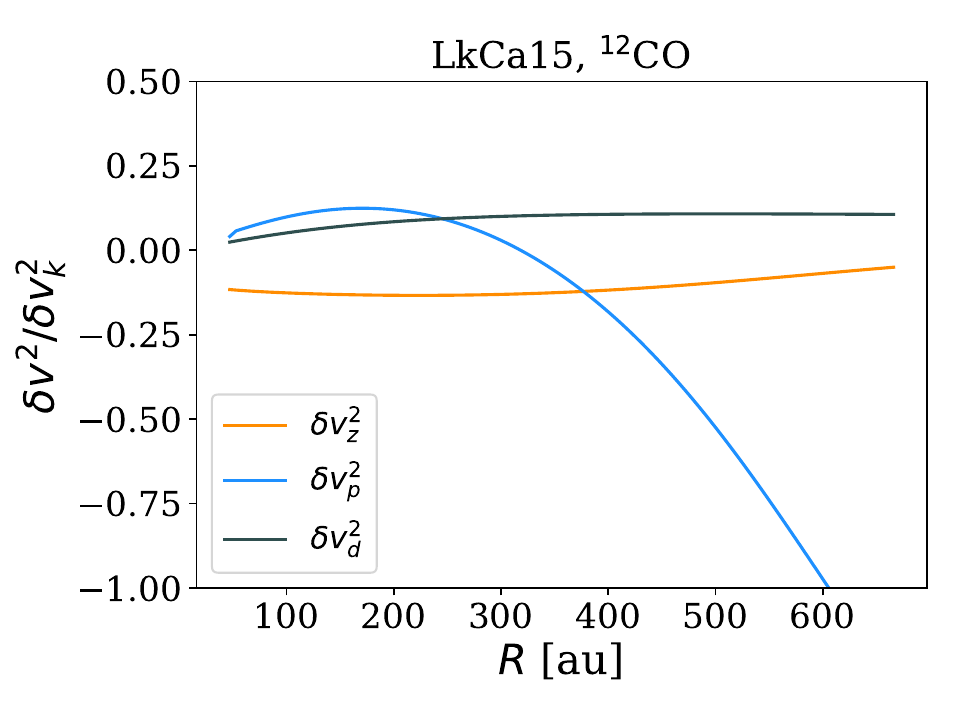}
    \caption{Top panels: rotation curves of LkCa15 (red dots) with the best fit model (blue lines) and residuals according to Eq. \eqref{rotationcurve_strat}. The black dashed line represents the location of the scale radius $R_c$. Bottom panel: Non-Keplerian contribution to the rotation curve, {where $\delta v_z$ is the correction due to the finite height of the emission, $\delta v_p$ is the pressure gradient and $\delta v_d$ is the self-gravitating contribution.}}
    \label{rotC_lkca}
\end{figure*} 

\begin{deluxetable*}{lccccccccc} \tablewidth{0pt} \tablecaption{Table summarizing the main results of this work. Best fit values for stellar mass, disk mass, and scale radius with the relative uncertainties from the posterior distributions, disk to star mass ratio with propagated errors, disk mass relative uncertainty, stellar mass from \textsc{discminer} fit \citep{Izquierdo_exoALMA}, percentage difference from the dynamical stellar mass estimate and the \textsc{discminer} one, dust mass from continuum emission from \cite{Curone_exoALMA}, and gas to dust ratio. The value $\sigma_{M_d}/M_d$ represents the average of asymmetric uncertainties (upper and lower) on the disk mass. The errors on the disk to star mass ratio and on the gas to dust ratio have been computed using the posterior distributions for the relevant quantities. \\ $\dagger$ These sources have a disk-to-star mass ratio $<0.05$, the theoretical limit for disk mass detection \citep{veronesi24,Andrews24}. We report the best-fit values obtained through the fitting procedure, regardless of the disk mass limit. In the figures, these sources are labeled with a diamond rather than a square. \label{table_results}} \tablehead{ \colhead{\textbf{Source}} & \colhead{$M_\star$ [M$_\odot$]} & \colhead{$M_{d}$ [M$_\odot$]} & \colhead{$R_c$ [au]} & \colhead{$M_d/M_\star$} & \colhead{$\sigma_{M_d}/M_d$} & \colhead{$M_{\rm discminer}$} & \colhead{$\Delta M_\star / M_\star$ [\%]} & \colhead{$M_{\rm dust}$ [M$_{\rm J}$]} & \colhead{$g/d$} } \startdata \textbf{AA Tau} & $0.624^{+0.033}_{-0.035}$ & $0.155^{+0.036}_{-0.036}$ & $156^{+106}_{-41}$ & $0.25^{+0.06}_{-0.06}$ & 0.23 & 0.791 & 26.7\% & 0.115 & $1417^{+326}_{-330}$ \\ \textbf{DM Tau} & $0.468^{+0.014}_{-0.015}$ & $0.057^{+0.019}_{-0.020}$ & $240^{+42}_{-27}$ & $0.12^{+0.04}_{-0.04}$ & 0.34 & 0.453 & -3.1\% & 0.162 & $367^{+131}_{-121}$ \\ \textbf{HD 34282} &$1.520^{+0.025}_{-0.031}$ & $0.143^{+0.045}_{-0.041}$ & $370^{+109}_{-78}$ & $0.09^{+0.03}_{-0.03}$ & 0.30 & 1.620 & 6.6\% & 1.091 & $137^{+40}_{-43}$ \\ \textbf{J1615} &$1.105^{+0.011}_{-0.012}$ & $0.082^{+0.014}_{-0.014}$ & $167^{+20}_{-15}$ & $0.07^{+0.01}_{-0.01}$ & 0.18 & 1.140 & 3.1\% & 0.308 & $279^{+49}_{-49}$ \\ \textbf{J1842} &$1.042^{+0.010}_{-0.011}$ & $0.078^{+0.013}_{-0.014}$ & $231^{+102}_{-50}$ & $0.07^{+0.01}_{-0.01}$ & 0.18 & 1.068 & 2.5\% & 0.108 & $759^{+134}_{-129}$ \\ \textbf{J1852}$^{\dagger}$ &$1.022^{+0.021}_{-0.021}$ & $0.044^{+0.024}_{-0.032}$ & $87^{+69}_{-16}$ & $0.04^{+0.02}_{-0.03}$ & 0.65 & 1.028 & 0.6\% & 0.110 & $420^{+300}_{-301}$ \\ \textbf{LkCa15} & $1.118^{+0.013}_{-0.015}$ & $0.108^{+0.016}_{-0.016}$ & $150^{+12}_{-16}$ & $0.10^{+0.01}_{-0.01}$ & 0.15 & 1.028 & -8.0\% & 0.333 & $339^{+49}_{-50}$ \\ \textbf{PDS66}$^{\dagger}$ & $1.299^{+0.036}_{-0.101}$ & $0.038^{+0.099}_{-0.035}$ & $28^{+12}_{-5}$ & $0.03^{+0.08}_{-0.03}$ & 1.76 & 1.277 & -1.7\% & 0.108 & $364^{+342}_{-357}$ \\ \textbf{SY Cha} &$0.812^{+0.037}_{-0.041}$ & $0.084^{+0.044}_{-0.043}$ & $112^{+21}_{-15}$ & $0.10^{+0.05}_{-0.05}$ & 0.52 & 0.813 & 0.1\% & 0.170 & $517^{+266}_{-273}$ \\ \textbf{V4046 Sgr}$^{\dagger}$ &$1.777^{+0.005}_{-0.006}$ & $0.058^{+0.006}_{-0.006}$ & $99^{+5}_{-5}$ & $0.03^{+0.00}_{-0.00}$ & 0.11 & 1.763 & -0.8\% & 0.112 & $540^{+60}_{-59}$ \\ \enddata \end{deluxetable*}

\subsection{Rotation curve fits}
The rotation curves used in this work have been extracted with \textsc{discminer} and thoroughly discussed in \cite{Stadler_exoALMA} using the default cubes. The emitting layers of $^{12}$CO and $^{13}$CO are an output of \textsc{discminer} as well. Finally the thermal structure are extracted with \textsc{disksurf} and thoroughly discussed in \cite{Galloway_exoALMA}. The thermal and geometrical parameters are summarized in Appendix \ref{appenidx_paramters}. For each disk, both the $^{12}$CO and $^{13}$CO rotation curves are fitted simultaneously.

The rotation curve fits were performed using the code \textsc{DySc}\footnote{\url{https://github.com/crislong/DySc}}, and the details can be found in Appendix \ref{appenidx_dysc}. All the \textsc{DySc} fits were performed using $10$ walkers, $5000$ steps of burnin and $5000$ steps. Initially, the walkers are uniformly distributed within their prior intervals. All the results are summarized in table \ref{table_results}, where we present the best fit values and their uncertainties, determined by the width of the posterior distributions. Hereafter, we will refer to the best fit parameters obtained through this methodology as ``dynamical values''. Figure \ref{rotC_lkca} shows the comparison between the data and the best fit model for the $^{12}$CO and $^{13}$CO rotation curves for LkCa15. The complete collection of comparison plots are shown in Appendix \ref{appenidx_bestfits}.  In our analysis we excluded the first two major beam sizes in radius, as discussed in \cite{Stadler_exoALMA}. The bottom panel of Figure \ref{rotC_lkca} shows the non-Keplerian contributions to the rotation curve predicted by the best fit model, where $\delta v_z$ is the contribution given by the finite height of the emitting layer, $\delta v_p$ is the pressure gradient contribution at the emitting layer and $\delta v_d$ is the self-gravitating contribution \citep{Lodato23}. 

When fitting AA Tau, we limited our analysis out to 250au for the $^{12}$CO, since the outer part of the disk shows a drop in the non-parametric emitting height extracted by $\textsc{disksurf}$, as shown in \cite{Galloway_exoALMA}. This feature is also visible in the channel maps and in the rotation curve (see Appendix \ref{appenidx_aatau}), and leads to a systematic shift of the best fit parameters, since our model can not take this into account. Despite excluding the outer region, the results for AA Tau still appear unusual, with the derived mass of the disk significantly high, marking AA Tau as an outlier in our sample. This is particularly clear when looking at the gas to dust ratio, at the percentage difference with the \textsc{discminer} stellar mass and at the $\alpha_S$. This disk presents several atypical characteristics, and its anomalous behavior could be attributed to its high inclination, affecting the extraction of the emitting height and rotation curve, as suggested in \cite{Galloway_exoALMA}. While we include the results for completeness, they should be interpreted with caution due to these reasons. Consequently, for the statistical analysis in this paper, we exclude AA Tau from the sample and represent it in the figures with a cross rather than a square to denote its peculiar status.

The issue of the diffuse backside is also observed in SY Cha at $R>400$ au and in the $^{13}$CO data of LkCa15 \citep{Galloway_exoALMA}. We tested the robustness of the fits by including and excluding these regions, demonstrating that the results remain consistent for these disks.

\subsection{Systematic uncertainties}

The \textsc{emcee} fitting procedure provides the uncertainties on the best fit parameters but does not account for systematic errors. Specifically, we model the pressure gradient contribution using the temperature structures presented in \cite{Galloway_exoALMA}. These temperature structures have their own uncertainties, which affect our results. To incorporate the systematic, we decided to run each \textsc{emcee} fit 100 times, drawing the thermal parameters from the posterior distribution of \cite{Galloway_exoALMA}. This procedure allows us to take into account the systematic errors driven by the thermal structure, returning a realistic uncertainty for the best fit parameters, listed in table \ref{table_results}.

In principle, the geometric parameters (inclination $i$ and position angle PA) and the height of the emitting layer can contribute to the systematic uncertainties. However, the thermal structure dominates the systematic uncertainties, since it  directly impacts the pressure gradient characterization. Additionally, as pointed out in \cite{Andrews24}, the position of the center of the disk may induce significant uncertainties; for more information on how the center of the images is found we refer to \cite{Izquierdo_exoALMA} for the gas and \cite{Curone_exoALMA} for the continuum.

\section{Results and discussion}\label{S4}

Before delving into the discussion of the results, we would like to emphasize that in Figures \ref{gtd_figure}, \ref{radii_comparison_plot}, \ref{trapman_radii}, and \ref{alpha_plot}, different markers are used: orange squares represent exoALMA sources with a disk-to-star mass ratio\footnote{A disk-to-star mass ratio of 5\% represents the lower limit for reliably measuring disk masses, as discussed by \cite{veronesi24} and \cite{Andrews24}.} $>$5\%, orange diamonds denote exoALMA sources with a disk-to-star mass ratio $<$5\%, the orange cross marks AA Tau, and blue squares indicate the MAPS sources. {The errors on MAPS sources are smaller compared to the exoALMA ones. This is because in \cite{martire24} the authors did not propagate the thermal structure uncertainties as we did in this work.}

\subsection{Stellar masses}
The best fit values for the stellar masses are summarized in table \ref{table_results}. The dynamical stellar masses generally deviate from the best fit values provided by \textsc{discminer} \citep{Teague_exoALMA,Izquierdo_exoALMA}. \textsc{discminer} operates under the assumption that the azimuthal velocity is solely dictated by the stellar gravity, disregarding the pressure gradient and disk self-gravity. Consequently, the dynamical masses obtained in this work are a more accurate estimate, as all the relevant non-Keplerian contributions are considered.

In general, the pressure gradient has a decelerating effect on the azimuthal velocity\footnote{The pressure gradient is always negative at the mid-plane, but this is not true at all $(R,z)$. Indeed, it depends on how the emitting layer $z(R)$ relates with the disk hydrostatic structure. For instance, in a vertically isothermal disk, for $z\gtrsim 2H$ the pressure gradient is positive rather than negative. For more details about the vertical isothermal model we refer to \cite{Lodato23}.}, leading to an underestimate of stellar mass when employing \textsc{discminer}. Conversely, the contribution from the disk self-gravity increases the azimuthal velocity, causing an overestimate the stellar mass. In table \ref{table_results} we observe that the combined effect of pressure gradient and self-gravity results in a discrepancy of few percents in the stellar masses compared to the simple Keplerian fit model, as employed by \textsc{discminer}, where we define $\Delta M = M_{\rm discminer} - M_{\rm dyn}$. In previous studies, dynamical masses were derived from low-resolution CO data and compared with Pre-Main-Sequence evolutionary track predictions \citep{simon00, simon17, simon19, braun21}. However, due to the limited resolution of these data, the rotational profiles were approximated as Keplerian. Thanks to the high spatial and spectral resolution exoALMA data, we can now model the rotational profile more precisely, resulting in more accurate stellar mass estimates.

In general, excluding AA Tau as discussed before, we note that the percentage difference between dynamical masses and the \textsc{discminer} ones is of the order of $\sim 5\%$. We also note that deviations go in both direction, showing that the combined contribution of self-gravity and pressure gradient is not easy to determine.%, and depends on the disk characteristics, specifically the functional form of the temperature structure and surface density profile.

%\cite{Rosenfeld2012} first studied arcsecond-resolution sub-millimetric observations of the $^{12}$CO J=2-1 line of the disk around the spectroscopic binary V 4046 Sgr. They obtained dynamical stellar mass measurement through CO kinematics, in very good agreement with the astrometric estimate $M_\star = 1.75^{+0.09}_{-0.06}\text{M}_\odot$. Our result is perfectly in line with their estimate. our result....

\subsection{Disk masses}
The best-fit values for disk masses are presented in Table \ref{table_results}. As outlined by \cite{veronesi24,Andrews24}, the minimum detectable disk-to-star mass ratio with this method is approximately 5\%. In our sample, only three sources —J1852, PDS66, and V4046— have best-fit disk mass values below this threshold. Table \ref{table_results} lists all best-fit values, regardless of the disk mass detection threshold. In the following figures, these sources are marked with diamonds instead of squares to indicate that they fall below the measurability threshold. One can decide whether to employ the best fit value we report, or to use $M_d=0.05M_\star$ as an upper limit. 

\begin{figure}
    \centering
    \includegraphics[width=\columnwidth]{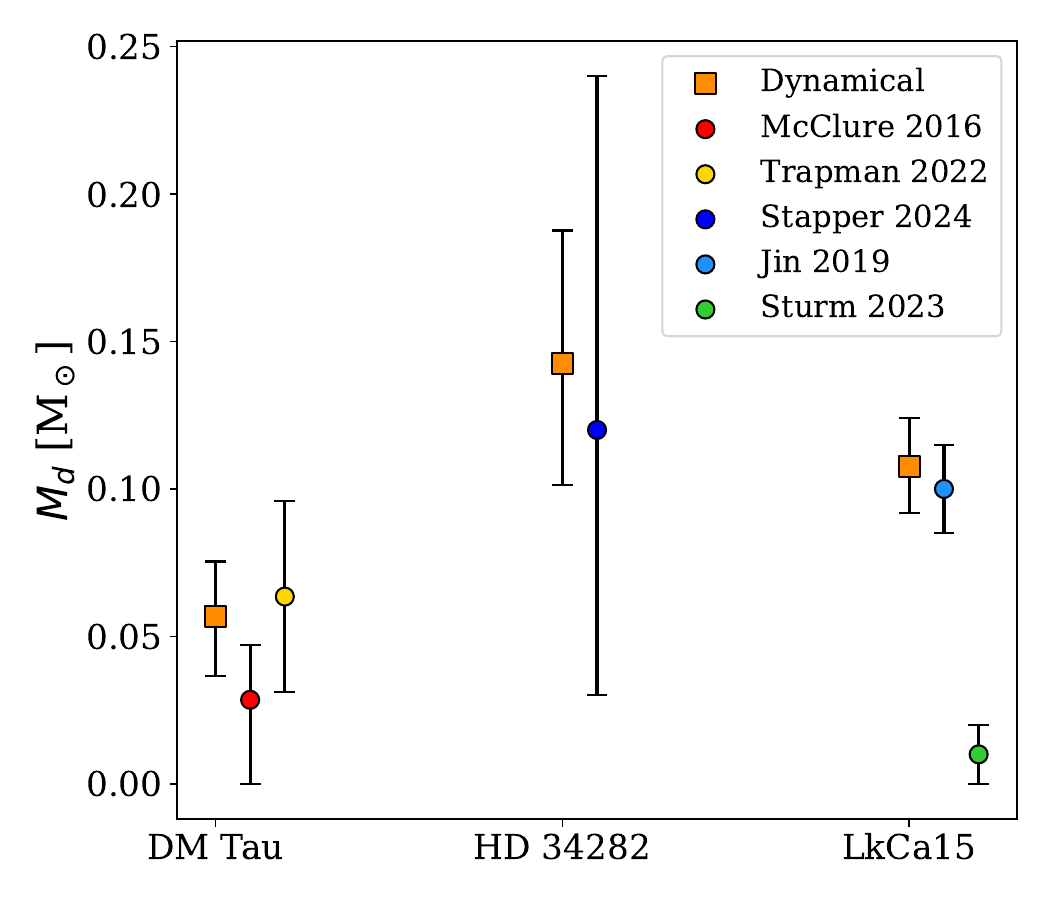}
    \caption{Comparison between dynamical disk masses (this work) and literature estimates. }
    \label{comparison_masses_plot}
\end{figure}

DM Tau, HD 34282, and LkCa 15 are the three sources whose disks have been independently estimated in the literature. In Figure \ref{comparison_masses_plot}, we compare the dynamical disk mass estimated in this paper to these independently estimates.

For DM Tau, we estimated a dynamical mass of $M_d =0.057^{+0.019}_{-0.020}{\rm M}_\odot$. Its mass was estimated from observation of hydrogen deuteride HD by \cite{mcclure16}. For a chosen disk model, they determined the disk mass to be between $0.01 \text{M}_\odot$ and $0.047 \text{M}_\odot$. Hydrogen deuteride is a powerful molecule to determine disk mass since there is no chemistry involved, being an isotopologue of molecular hydrogen. \cite{trapman22} presented an innovative approach to measure protoplanetary disk masses using N$_2$H$^+$ and C$^{18}$O. This method enables the determination of the CO-to-H$_2$ mass ratio, facilitating the calibration of CO-based mass measurements \citep[see][for recent reviews]{Oberg23,Miotello23}. They applied this technique to DM Tau, determining that the disk mass lies between $0.031 \text{M}_\odot$ and $0.096 \text{M}_\odot$. To compare with the HD based measurements of the disk mass, they repeated the analysis of \cite{mcclure16}, but  assuming a different disk structure, consistent with N$_2$H$^+$ models. They found a slightly higher HD based disk mass between $0.04 \text{M}_\odot$ and $0.2 \text{M}_\odot$, being the upper limit at the edge of gravitational instability. The value we obtained in our study is consistent with both N$_2$H$^+$ and HD based measurements of \cite{trapman22}.

For HD 34282 we estimated a dynamical mass of $M_d =0.143^{+0.045}_{-0.041}{\rm M}_\odot$. \cite{stapper24} constrained gas masses of Herbig disks using CO isotopologues, and HD 34282 is within their sample. Their estimate is $M_d = 0.12^{+0.19}_{-0.09}\text{M}_\odot$, perfectly consistent with our dynamical value.

For LkCa 15 we estimated a dynamical mass of $M_d =0.108^{+0.016}_{-0.016}{\rm M}_\odot$. \cite{jin19} constrain gas and dust distribution in the LkCa15 disk by comparing radiative transfer models with the $^{12}$CO and dust emission from ALMA. They found that the best fit value for the disk mass is $M_d=0.1\text{M}_\odot$, that reproduces very well gas and dust emission. Conversely, \cite{sturm23} measured LkCa15 gas mass using CO, $^{13}$CO, C$^{18}$O and C$^{17}$O lines and modeling them with thermochemical models. They found that the gas mass in the LkCa15 disk is $M_d=0.01\rm M_\odot$, an order of magnitude smaller compared to \cite{jin19} estimate. The main differences between the two methods lie in the assumptions about the disk structure, in particular in the disk surface density profile. Our estimate is consistent with \cite{jin19} and $\gtrsim 5 \sigma$ inconsistent with \cite{sturm23}.

PDS 66, or MP Mus, does not exhibit any substructure in dust continuum emission or gas, even at very high angular resolution $\Delta x = 0.05^{\prime \prime}$ \citep{ribas23}, and in the rotation curves analyzed here there are no clear signs of pressure-induced substructures \citep{stapper24}. \cite{ribas23} estimated the disk mass of PDS 66 comparing the CO isotopologues' fluxes with models of \cite{williams14} and \cite{miotello16}, showing that $M_d \simeq 10^{-4}-10^{-3}\text{M}_\odot$. Their estimate is consistent with our results, assuming $M_d/M_\star<0.05$.

V4046 Sgr  is a known spectroscopic binary with a period of 2.4 days and a mass ratio $q\simeq0.94$ \citep{stempels04}. \cite{miotello16} presented chemical models to infer the disk mass from CO isotopologues emission. In that work, they gave an estimate of V4046 Sgr disk mass being $\sim10^{-3}\rm M_\odot$. The result is consistent when we assume the $M_d<0.05M_\star$ upper limit.

A comparison between dynamical masses and chemical masses obtained through the modeling of the N$_2$H$^+$ emission for the exoALMA sample is presented in \cite{Trapman_exoALMA}.

\subsubsection{Gas to dust ratio}
\begin{figure}
    \centering
    \includegraphics[width=\linewidth]{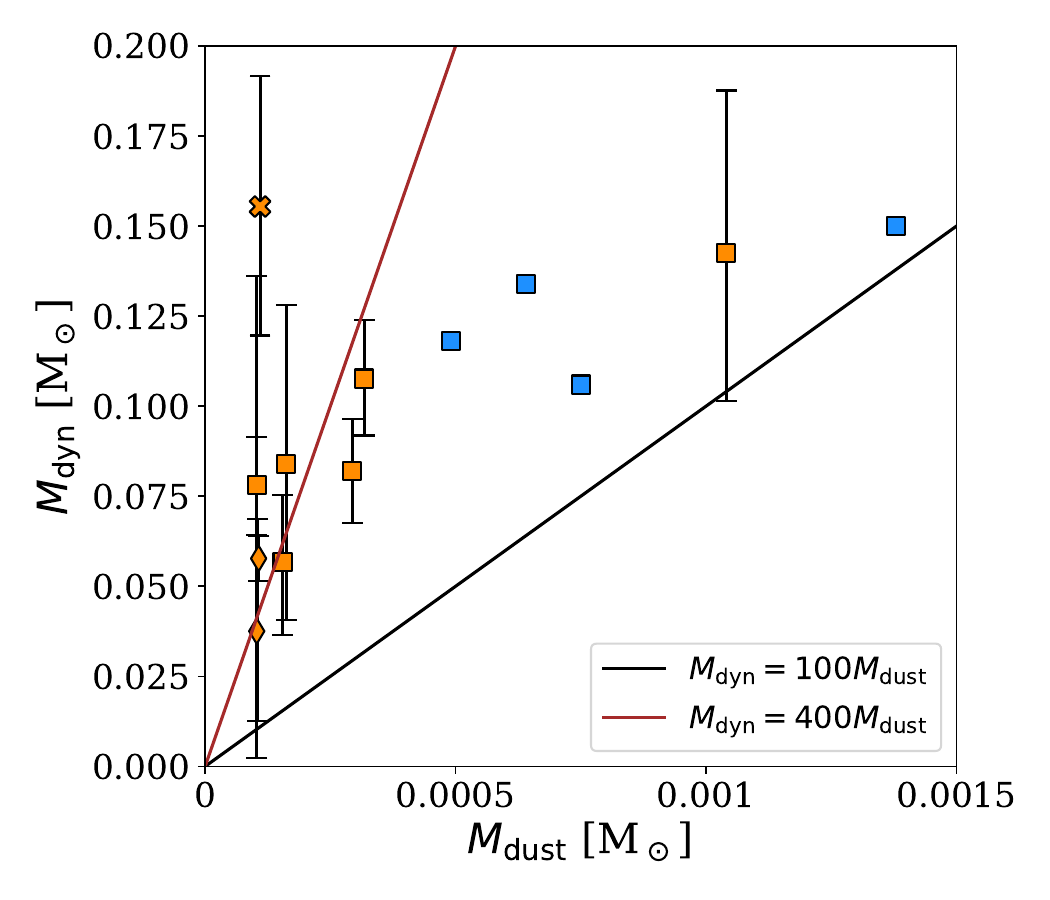}
    \caption{{Dynamical masses against dust masses as computed in \cite{Curone_exoALMA} for the exoALMA and MAPS sources. The black line shows the $M_{\rm dyn}=100M_{\rm dust}$ and the brown line the $M_{\rm dyn}=400M_{\rm dust}$.}}
    \label{gtd_figure}
\end{figure}

We use the fiducial values for the disk masses to compute the gas-to-dust ratio,  employing the dust masses obtained by \cite{Curone_exoALMA}. {Table \ref{table_results} presents the gas-to-dust ratios for the disks within the exoALMA sample, and Figure \ref{gtd_figure} shows the dynamical masses compared to the dust ones.} The sources indicated with a diamond are the one below the minimum disk mass threshold. Overall, the gas-to-dust ratios are above the standard value of 100, with an average value of $\sim 400$. To compute the average gas-to-dust ratio, we have excluded AA Tau. It is not surprising that the inferred values are above 100: indeed, dust masses computed in \cite{Curone_exoALMA} underestimate the total dust mass, because of the optically thin emission hypothesis. Indeed, we expect the sources within the sample to be, at least, marginally optically thick in the inner parts. In addition, \cite{stapper24} estimated gas masses of Herbig disks, and compared them with dust masses to obtain the gas to dust ratio. In their sample, they also observes a mean value of the gas to dust ratio $\sim 400$, as we find in this work. {Interestingly, Fig. \ref{gtd_figure} suggests that the gas-to-dust ratio is higher in low dust mass disks. These disks also tend to be more compact, leading to higher optical depths, as it scales with the surface density.}

\subsubsection{Gravitational instability?}
\begin{figure*}
    \centering
    \includegraphics[width=0.85\textwidth]{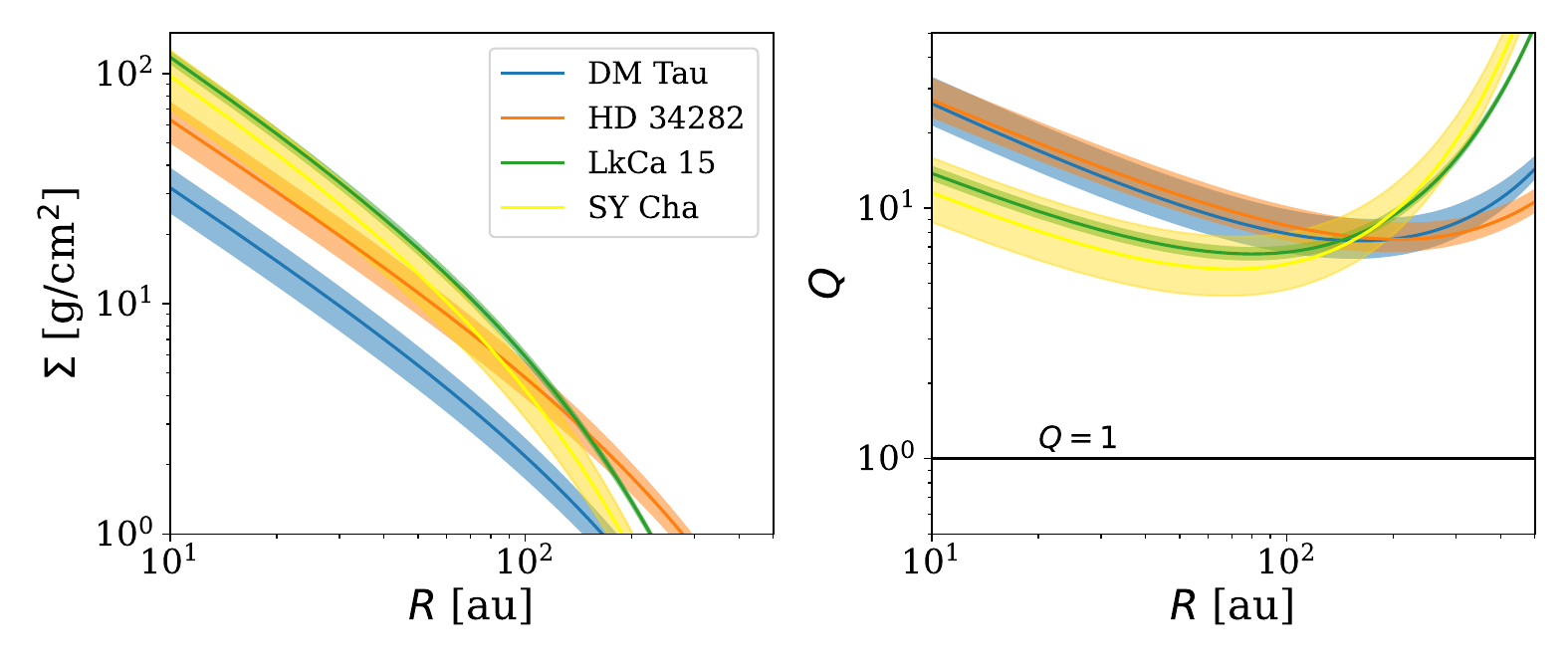}
    \includegraphics[width=0.85\textwidth]{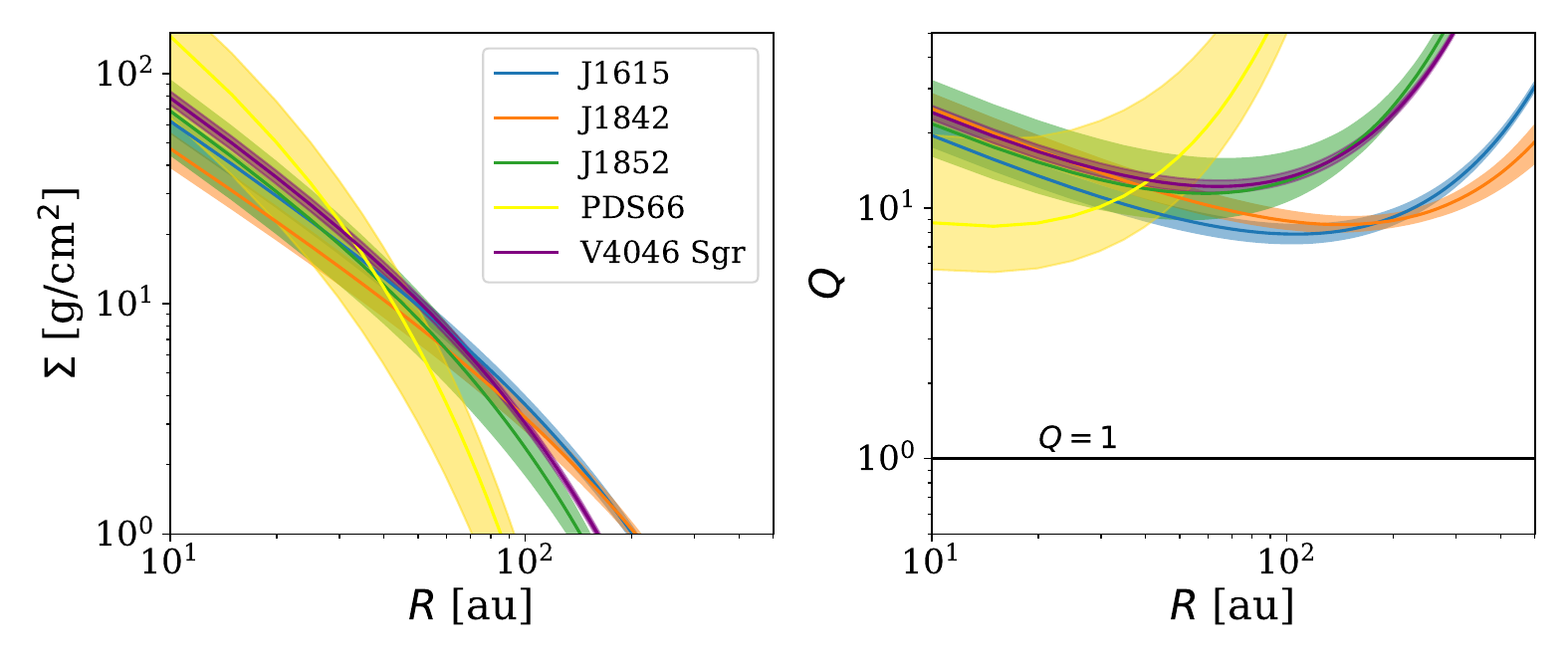}
    \caption{Top panel: Surface density and Toomre Q profiles for the four most massive disks of our sample, namely DM Tau, HD 34282, LkCa 15 and SY Cha, excluding AA Tau because of the big uncertainties (see Appendix \ref{appenidx_aatau}).
    Bottom panel: Surface density and Toomre Q profiles for the other 5 sources, namely J1615, J1842, J1852, PDS66 and V4046 Sgr.}
    \label{ToomreQ_plot}
\end{figure*}

To investigate the likelihood of a disk to be gravitationally unstable, we compute the Toomre Q parameter, defined as \citep{toomre64}
\begin{equation}
        Q \simeq \frac{c_{s,\text{mid}}\Omega_\text{k}}{\pi G \Sigma} = 2\left.\frac{H}{R}\right|_\text{mid}\left(\frac{M_\star}{M_d}\right) \left(\frac{R}{R_c}\right)^{-1}\exp\left[\frac{R}{R_c}\right].
\end{equation}
This dimensionless parameter measures the strength of the stabilizing terms, pressure $(c_s)$ and rotation $(\Omega_\text{k})$ against the disk self gravity $(\Sigma)$. According to the WKB (Wentzel–Kramers–Brillouin) quadratic dispersion relation \citep{linshu64,toomre64}, the onset of the instability occur when $Q \sim 1$. Figure \ref{ToomreQ_plot} shows the surface density profiles and the Toomre parameter profiles for the disks within our sample. We are excluding AA Tau because of the large uncertainties, and its case is thoroughly commented in Appendix \ref{appenidx_aatau}. The massive disks in our sample show $Q>1$, meaning that they all are gravitationally stable. We underline that the temperature at the midplane is extrapolated from the 2D thermal structures \citep{Galloway_exoALMA}.

\subsection{Scale radii}
In this section we discuss the relationship between the flux based radii, i.e. the radii enclosing the 68\% of the $^{12}$CO, $^{13}$CO \citep{Galloway_exoALMA} and dust emission \citep{Curone_exoALMA}, and the scale radii $R_c$ we find  by modeling the rotation curve. As part of the discussion, we also add the MAPS sources. The masses and scale radii of the MAPS sources are taken from \cite{martire24}, and the flux based radii from \cite{MAPSIV}.

\subsubsection{Gas based measurement}\label{radii_gas}
\begin{figure}
    \centering
    \includegraphics[width = \columnwidth]{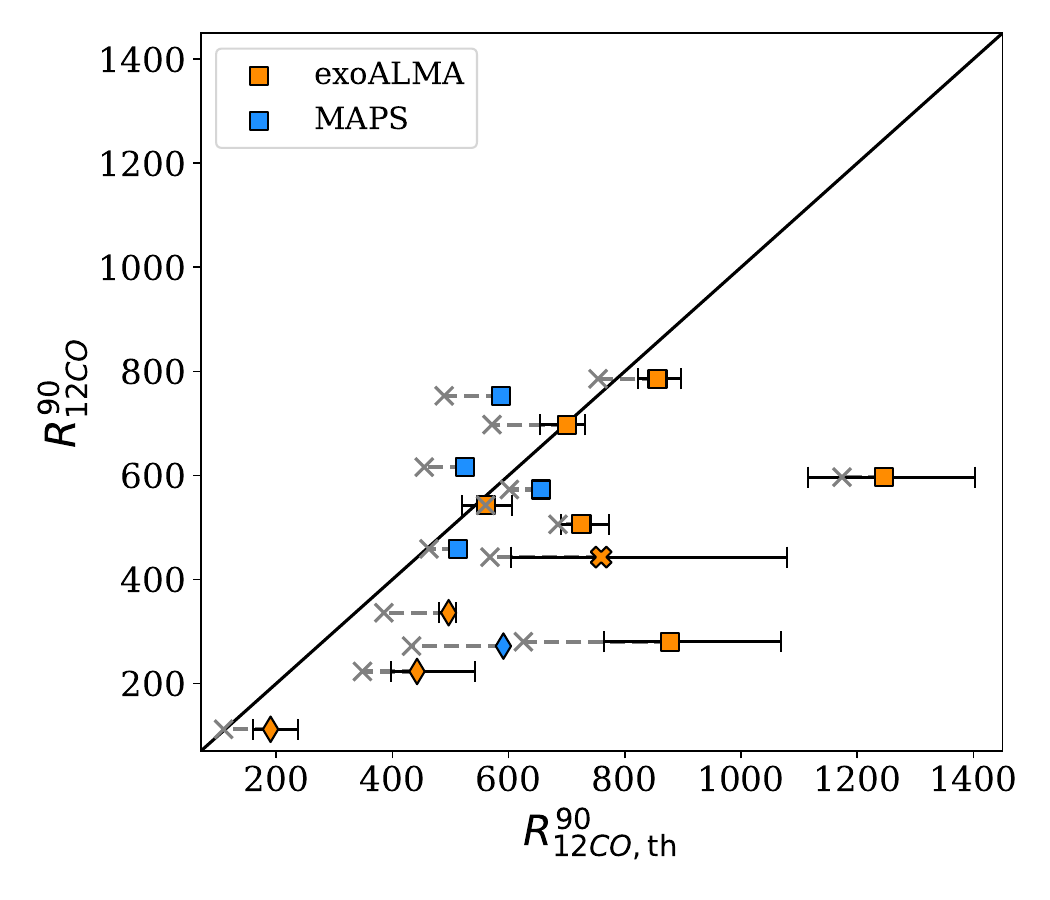}
    \caption{Comparison between the observed and the predicted radius enclosing the 90\% of the $^{12}$CO emission, according to Eq. \eqref{trap_rad}. The grey crosses show the results assuming the CO depletion obtained by \cite{Trapman_exoALMA}.}
    \label{trapman_radii}
\end{figure}

The left and the central panels of Figure \ref{radii_comparison_plot} display the comparison between the dynamical and the flux based radii  $^{12}$CO and $^{13}$CO. As expected, the flux-based radii are larger compared to the dynamical scale radius, showing an average ratio of $2.5$ for the $^{12}$CO and $1.75$ for the $^{13}$CO. However, $R_c$ is a crucial quantity in the context of protoplanetary disk evolution, as it relates with the disk's lifetime and the efficiency of radial drift. The relationship between the flux-based radii and $R_c$ is complex, involving thermochemical information and potentially depending on the sensitivity of the observations. This issue was addressed by \cite{toci23,trapman23}, who found a relationship between $R_{12\rm CO}^{90}$ and $R_c$ and $M_d$, which reads
\begin{equation}\label{trap_rad}
    R_{12\rm CO, theor}^{90} = R_c \mathcal{W}\left[4.9\times10^7 \left(\frac{M_d}{M_\odot}\right)^{0.66} \left(\frac{R_c}{\rm au}\right)^{-2}\right],
\end{equation}
where $\mathcal{W}$ is the Lambert function. This expression assumes a CO abundance, that typically is considered to be $X_{\rm CO}= 10^{-4}$, and this information is enclosed into the constant $4.9\times10^{7}$. Figure \ref{trapman_radii} shows the comparison between the radius enclosing the 90\% of the $^{12}$CO emission and the theoretical expectation according to Eq. \eqref{trap_rad}. The error bars take into account the uncertainty on the disk mass and scale radius. We observe that the theoretical expectations systematically overestimate the CO radius. A possible explanation is CO depletion. \cite{trapman23} obtain Eq. \eqref{trap_rad} using thermochemical models, fixing the CO abundance at $X_{\rm CO}=10^{-4}$. There, the dependence on CO abundance is not explicitly stated. Assuming a linear relationship between the CO abundance and the argument of the Lambert function in Eq. \eqref{trap_rad}, as done in \cite{toci23}, we see that reducing the CO abundance brings the theoretical values closer to the observed ones. \cite{Rosotti_exoALMA} and \cite{Trapman_exoALMA} provided two distinct inferences of CO depletion for the exoALMA sample. \cite{Rosotti_exoALMA} derived the CO depletion required to reconcile their estimated disk masses with the dynamical values presented in this paper, meaning their method is not entirely independent of our estimates. In contrast, \cite{Trapman_exoALMA} presented an independent estimate of CO depletion by forward-modeling N$_2$H$^+$ and rare CO isotopologue emission in the exoALMA disks. Both studies consistently indicate that CO depletion is needed, aligning with the discrepancy we report. The grey crosses in Fig. \ref{radii_comparison_plot} represent the values of $R_{\rm CO}$ calculated using the CO abundance derived by \cite{Trapman_exoALMA}. For most of the sources, using these CO abundances result in a better agreement with the observed values.

\subsubsection{Dust based measurement}
\begin{figure*}
    \centering
    \includegraphics[width = \textwidth]{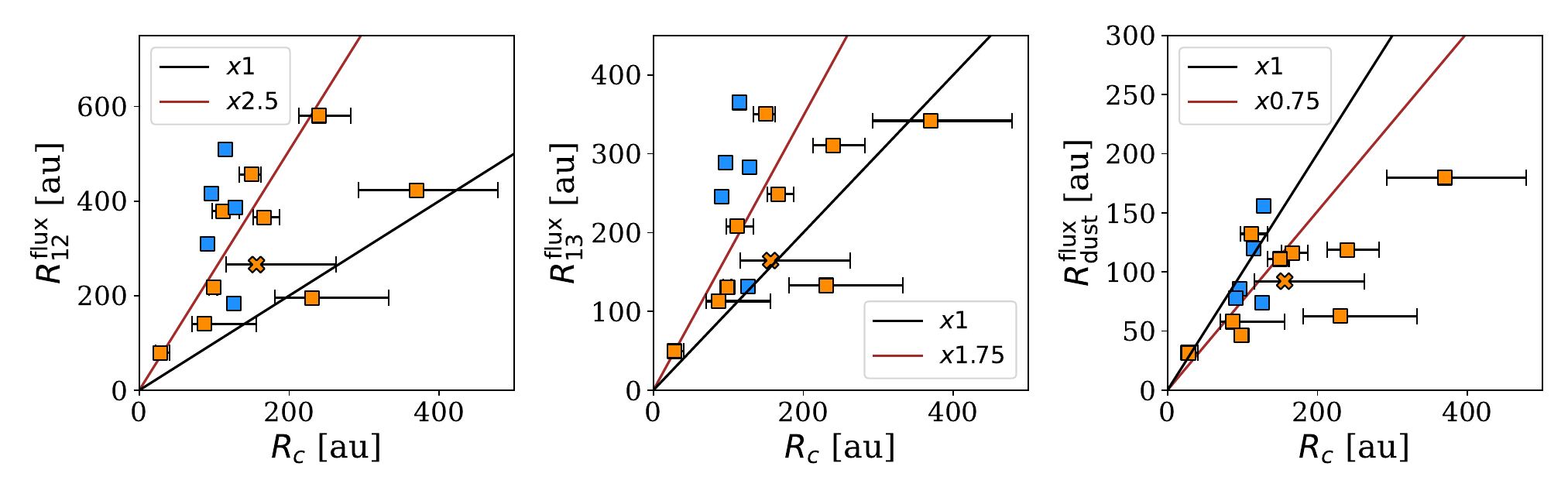}
    \caption{Flux based radii (i.e. radii enclosing the 68\% of the emission) of $^{12}$CO, $^{13}$CO and dust compared with the dynamical scale radii $R_c$. The orange squares are the exoALMA sources, while the blue ones are the MAPS. The black line shows when the flux radius is equal to the dynamical one, and the brown line is the average value for the sources. }
    \label{radii_comparison_plot}
\end{figure*}

The right panel of Figure \ref{radii_comparison_plot} displays the comparison between the dynamical and the flux based radius of dust emission. On average, the dust radii are smaller than the scale radii, showing an average ratio of $0.75$. This trend is expected because, beyond $R_c$, the surface density profile is exponentially tapered, enhancing the effect of radial drift \citep{birnstiel14}. 

\cite{toci21} presented theoretical models of protoplanetary disk evolution influenced by viscosity, grain growth, and radial drift, and studied the ratio between the scale radius and the dust. They found that the expected ratio between the scale and dust radii should be $\gtrsim 5$, significantly larger than the average ratios measured in our samples. One possible reason may be the role of substructures, that slow down radial drift. Indeed, all the disk, except for PDS66, show substructures in dust continuum emission; for a detailed characterization of dust substructures, we refer to \cite{Curone_exoALMA}.

\begin{deluxetable}{lccc}
\tablewidth{0pt}
\tablecaption{Flux-based radii of dust, $^{12}$CO, and $^{13}$CO emission for the exoALMA and MAPS sources \citep{Galloway_exoALMA,MAPSIV}. \label{table_flux_radii}}
\tablehead{
\colhead{\textbf{Source}} & 
\colhead{$R_{d}^{68}$ [au]} & 
\colhead{$R_{\rm 12CO}^{68}$ [au]} & 
\colhead{$R_{\rm 13CO}^{68}$ [au]}
}
\startdata
\textbf{AA Tau} & 92  & 265  & 164 \\
\textbf{DM Tau} & 119 & 580  & 310 \\
\textbf{HD 34282} & 180 & 422  & 341 \\
\textbf{J1615} & 116 & 365  & 248 \\
\textbf{J1842} & 63  & 195  & 133 \\
\textbf{J1852} & 58  & 140  & 113 \\
\textbf{LkCa15} & 111 & 457  & 351 \\
\textbf{PDS66} & 32  & 108  & 50 \\
\textbf{SY Cha} & 132 & 376  & 207 \\
\textbf{V4046 Sgr} & 46  & 216  & 130 \\
\\
\textbf{AS 209} & 74  & 184  & 132 \\
\textbf{GM Aur} & 86  & 416  & 289 \\
\textbf{HD 163296} & 78  & 310  & 246 \\
\textbf{IM Lup} & 120 & 509  & 365 \\
\textbf{MWC480} & 156 & 387  & 283 \\
\enddata
\end{deluxetable}

\subsection{Transport of angular momentum - effective $\alpha_S$}
\begin{deluxetable*}{lccccc}
\tablecaption{Accretion rate and $\alpha_S$ for the exoALMA and MAPS sources. The literature values for $\alpha_S$ are based on different methods, mainly line broadening. \label{alpha_table}}
\tablewidth{0pt}
\tablehead{
\colhead{\textbf{Source}} & 
\colhead{$\log_{10}\dot{M}_\star$ [M$_\odot$/yr]} & 
\colhead{$\dot{M_\star}$ Reference} & 
\colhead{$\alpha_S$} & 
\colhead{Literature values} & 
\colhead{Reference}
}
\startdata
\textbf{AA Tau} & $-7.35 \pm 0.35$ & \cite{manara23} & $6.54^{+8.11}_{-3.62}\times10^{-3}$ & ... & ... \\
\textbf{DM Tau} & $-8.2 \pm 0.35$ & \cite{manara14} & $2.19^{+2.72}_{-1.21}\times10^{-3}$ & $0.08 \pm 0.02$ & \cite{flaherty20} \\
\textbf{HD 34282} & $-7.69 \pm 0.35$ & \cite{fairlamb15} & $3.86^{+4.79}_{-2.14}\times10^{-3}$ & ... & ... \\
\textbf{J1615} & $-8.25 \pm 0.35$ & \cite{manara14} & $1.20^{+1.48}_{-0.66}\times10^{-3}$ & ... & ... \\
\textbf{J1842} & $-8.8 \pm 0.35$ & \cite{manara14} & $4.03^{+4.99}_{-2.23}\times10^{-4}$ & ... & ... \\
\textbf{J1852} & $-8.7 \pm 0.35$ & \cite{manara14} & $3.62^{+4.49}_{-2.01}\times10^{-4}$ & ... & ... \\
\textbf{LkCa15} & $-8.7 \pm 0.35$ & \cite{manara14} & $1.74^{+2.16}_{-0.97}\times10^{-4}$ & ... & ... \\
\textbf{PDS66} & $-9.18 \pm 0.35$ & \cite{ingleby13} & $8.06^{+9.99}_{-4.46}\times10^{-5}$ & ... & ... \\
\textbf{SY Cha} & $-9.89 \pm 0.35$ & \cite{manara23} & $1.66^{+2.05}_{-0.92}\times10^{-5}$ & ... & ... \\
\textbf{V4046 Sgr} & $-9.3 \pm 0.35$ & \cite{donati11} & $1.11^{+1.38}_{-0.61}\times10^{-4}$ & $<0.014$ & \cite{flaherty20} \\ \\ 
\textbf{AS 209} & $-7.3 \pm 0.35$ & \cite{oberg21} & $>1.11\times10^{-2}$ & ... & ... \\
\textbf{GM Aur} & $-8.1 \pm 0.35$ & \cite{oberg21} & $9.32^{+11.1}_{-5.16}\times10^{-4}$ & ... & ... \\
\textbf{HD 163296} & $-7.4 \pm 0.35$ & \cite{oberg21} & $4.31^{+5.35}_{-2.38}\times10^{-3}$ & $<0.003$ & \cite{flaherty15, flaherty17} \\
\textbf{IM Lup} & $-7.9 \pm 0.35$ & \cite{oberg21} & $1.48^{+1.18}_{-0.82}\times10^{-3}$ & $3.0^{+0.4}_{-0.9}\times10^{-3}$ & \cite{franceschi23} \\
 & & & & $0.25^{+0.09}_{-0.09}$ & \cite{paneque23} \\
 & & & & $5.76_{-2.52}^{+8.68}\times10^{-2}$ & \cite{flaherty24} \\
\textbf{MWC 480} & $-6.9 \pm 0.35$ & \cite{oberg21} & $1.41^{+1.17}_{-0.78}\times10^{-2}$ & $<0.006$ & \cite{flaherty20} \\
\enddata
\end{deluxetable*}

We modeled the disks in the exoALMA sample under the assumption of a self-similar surface density distribution. This approach allowed us to fit the stellar mass, disk mass, and scale radius using the rotation curves of CO isotopologues. Consequently, we now have a comprehensive picture of the disks' structure, enabling us to study their evolution within a viscous framework \citep{shakurasun73}, introducing an effective viscous parameter $\alpha_S$. Indeed, the self-similar hypothesis establishes relationships between disk properties such as temperature, mass, size, and  $\alpha_S$ with the accretion rate onto the central object. By measuring the accretion rates, we can in principle constrain the instantaneous $\alpha_S$ using (i.e. \cite{hartmann98})
\begin{equation}\label{alpha_mdot}
    \alpha_S = \frac{2}{3} \frac{\dot{M}_\star}{M_d\Omega_c}\left(\frac{H_{c}}{R_{c}}\right)^{-2},
\end{equation}
where the subscript $c$ denotes that the corresponding quantity is evaluated at the scale radius.

The reader should remember that $\alpha_S$ was introduced to explain why disks accrete, and it is not a viscosity theory. Therefore, the correct interpretation of the formula above is the effective value of $\alpha_S$ needed to reproduce the observed accretion rate. Additionally, the formula does not imply that accretion is driven by turbulence; the values it returns should be interpreted as an effective $\alpha_S$, i.e. the amount of transported angular momentum. Any mechanism proposed to explain angular momentum transport in disks would need to exhibit an equivalent efficiency of angular momentum transport, even MHD winds \citep{tabone22}. We point out that here we are assuming that stellar accretion, which happens in the $<1$au region of the disk is equal to the  disk accretion, measured at the scale radius.

Several studies tackled this problem in the past \citep{andrews09,andrews10,rafikov17,ansdell18,vandermarel21}, showing that $\alpha_S > 10^{-4}$ are needed to explain the observed accretion rate. However, disk masses were estimated through dust emission, and the scale radius was modeled from dust emission or interpreted as a fixed fraction of the $^{12}$CO spectral line flux.  Here, we are able to correctly determine the $\alpha_S$ needed to explain accretion, since we have a dynamical disk mass estimate and a good measurement of the scale radius. 

\begin{figure}
    \centering
    \includegraphics[width=\columnwidth]{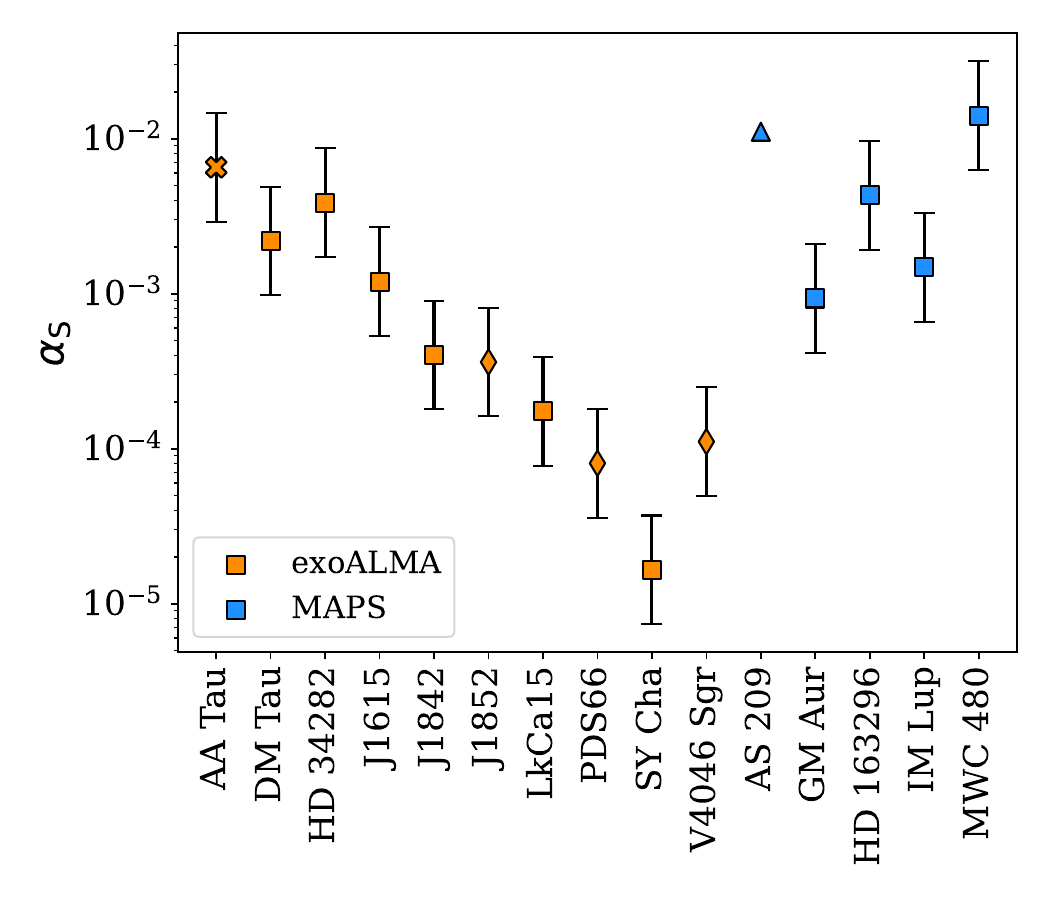}
    \caption{$\alpha_S$ for the exoALMA and MAPS sources, computed according to Eq. \ref{alpha_mdot} and comparison with literature values. }
    \label{alpha_plot}
\end{figure}

Figure \ref{alpha_plot} presents the values of $\alpha_S$ calculated using equation \eqref{alpha_mdot} for the exoALMA and MAPS sources. In the exoALMA sample, the diamonds represent the disc with a disk mass below measurable threshold, and for AS 209 we use $M_d=0.05M_\star$ as an upper limit, as commented in \cite{martire24}. Table \ref{alpha_table} provides the detailed values along with their associated uncertainties. The determination of $\alpha_S$ involves two primary sources of error: the uncertainties related to the accretion rate and the parameters of the disk. Our analysis indicates that the uncertainties in $\alpha_S$ are predominantly influenced by the uncertainties in the accretion rate, $\dot{M}_{\star}$. As reported by \cite{manara23}, the fractional uncertainty in individual accretion rate measurements at any given time is approximately 0.35 dex, a value which we have adopted for our analysis. Overall, the effective $\alpha_S$ we obtain is $> 10^{-5}$ in all cases, and $>10^{-4}$ in most cases. What is peculiar in Figure \ref{alpha_plot} is that the range of $\alpha_S$ is broad,  from $10^{-5}$ to $10^{-2}$, possibly pointing to different mechanisms driving angular momentum transport, or even accretion rate variability.

For five sources within our sample, namely DM Tau, V4046 Sgr, MWC480, IM Lup and HD163296, there are independent $\alpha_S$ constraints, obtained by modeling the non-thermal molecular line broadening \citep{flaherty15,flaherty17,flaherty20} or by determining the dust emission radial profile with radiative transfer models \citep{franceschi23}. The literature values are listed in table \ref{alpha_table}.

The case of DM Tau is particularly interesting since the two estimates disagree by 2 orders of magnitude, with our value pointing towards a lower viscosity $\alpha\sim10^{-3}$. To model the non-thermal line broadening in DM Tau, \cite{flaherty20} assume a stellar mass of $M_\star = 0.54\text{M}_\odot$, that differs of $\sim 20\%$ from our best fit value $M_\star = 0.456\text{M}_\odot$. This difference in stellar mass could explain the inconsistency between the $\alpha-$values: indeed, the lower the stellar mass, the bigger is the iso-velocity region for a fixed velocity interval $\Delta v$.

As for V4046 Sgr, our lower value is in agreement with \cite{flaherty20} upper value, pointing to $\alpha \in (10^{-4},10^{-2})$. For the MAPS sources MWC480, and HD163296, our estimate is in overall good agreement with \cite{flaherty17,flaherty20} upper limits. Finally, for IM Lup \cite{franceschi23} modeled dust emission with different $\alpha_S$ values, post-processed the models with radiative transfer codes and compared with the actual data. They found that an $\alpha_S$ of $3\times10^{-3}$ best reproduces the data. Their estimate agrees with our value within the uncertainties. By contrast, \cite{paneque23} estimate the viscosity through the characterization of CN and C$_2$H. Leveraging on the optical depth properties of these two tracers, they measured the turbulence from the non-thermal broadening of the line, at the location of the emitting layer. Recently, also \cite{flaherty24} estimated the $\alpha$ for IM Lup by molecular line broadening model, showing a good agreement with \cite{paneque23}. They found a high value of viscosity, almost two order of magnitude higher compared to our estimate and \cite{franceschi23} one, pointing to a vertical gradient of $\alpha$, as expected from instabilities like the MRI.

\section{Conclusions}\label{conclusions}
High-resolution rotation curves of protoplanetary disks can be used to constrain fundamental disk properties, namely the stellar mass, the disk mass, and the scale radius. In this work, we analyzed rotation curves of $^{12}$CO and $^{13}$CO for the exoALMA sources to infer these parameters. Here we summarize our findings

\begin{itemize}

    \item We constrained the dynamical disk mass for 10 sources within the exoALMA sample. Combined with the results from \cite{martire24} and \cite{veronesi21}, this brings the total number of dynamical disc mass estimates to 16. This method is independent of assumptions about disc chemical composition and does not rely on any specific tracer. Among the 10 sources analyzed in the exoALMA sample, 7 exhibit a disc-to-star mass ratio exceeding 5\%. We evaluated the Toomre parameter to assess gravitational stability and found that all sources are gravitationally stable, consistent with the absence of prominent spiral structures.

    \item We compared the dynamical disk masses with the dust based ones \citep{Curone_exoALMA}, with the assumption of optically thin continuum  emission, to determine the gas to dust ratio. We found values consistently above the standard  100, with an average of approximately 400. These large ratios likely result from the underestimation of dust masses due to the assumption of optically thin emission.

    \item Thoroughly modeling the pressure gradient contribution allows for accurate estimation of the scale radius $R_c$. We compared the scale radius estimates with flux-based measurements for CO isotopologues and dust, finding that the dust continuum emission radii are comparable to $R_c$. This suggests that pressure-modulated substructures may mitigate radial drift. Additionally, we find that the gas based radii are consistently larger than $R_c$. Using the derived $R_c$ and $M_d$, we calculated the theoretical flux based CO radii \citep{trapman23} and compared with the observed values. The theoretical predictions systematically overestimate the CO radii, possibly indicating CO depletion. We recomputed the CO radii using the CO depletion factor derived by \cite{Trapman_exoALMA} through forward-modeling N$_2$H$^+$ and C$^{18}$O  emissions. For most of the sources, this results in a better agreement.

    \item Correctly modeling the non-Keplerian contributions to the rotation curves allows for precise estimates of stellar masses. The dynamical stellar masses, which incorporate the effects of pressure gradients and disk self-gravity, provide a more accurate estimate of this quantity compared to simple Keplerian models, as shown in \cite{martire24} and \cite{Andrews24}.

    \item The knowledge of $M_d$, $M_\star$ and $R_c$ allows us to investigate protoplanetary disc evolution, particularly the transport of angular momentum within an $\alpha_S$ description. Our results show that the effective $\alpha_S$ for the disks is generally $> 10^{-5}$, with statistical uncertainties driven primarily by the accretion rate measurements.

\end{itemize}

%This work underscores the importance and the challenges of constraining fundamental parameters of protoplanetary disks, namely $M_\star$, $M_d$, and $R_c$.% Accurate measurements of these parameters are essential for advancing our understanding of protoplanetary disk dynamics and evolution.

\section*{Acknowledgments}
The authors thank the referee for the insightful report that significantly improved the quality of the paper.

This paper makes use of the following ALMA data: ADS/JAO.ALMA\#2021.1.01123.L. ALMA is a partnership of ESO (representing its member states), NSF (USA) and NINS (Japan), together with NRC (Canada), MOST and ASIAA (Taiwan), and KASI (Republic of Korea), in cooperation with the Republic of Chile. The Joint ALMA Observatory is operated by ESO, AUI/NRAO and NAOJ. The National Radio Astronomy Observatory is a facility of the National Science Foundation operated under cooperative agreement by Associated Universities, Inc. We thank the North American ALMA Science Center (NAASC) for their generous support including providing computing facilities and financial support for student attendance at workshops and publications. CL thanks Francesco Zagaria, Álvaro Ribas, Cathie Clarke, Leon Trapman, Claudia Toci.JB acknowledges support from NASA XRP grant No. 80NSSC23K1312. MB, DF, JS have received funding from the European Research Council (ERC) under the European Union’s Horizon 2020 research and innovation programme (PROTOPLANETS, grant agreement No. 101002188). Computations by JS have been performed on the `Mesocentre SIGAMM' machine,hosted by Observatoire de la Cote d’Azur. PC acknowledges support by the Italian Ministero dell'Istruzione, Universit\`a e Ricerca through the grant Progetti Premiali 2012 – iALMA (CUP C52I13000140001) and by the ANID BASAL project FB210003. SF is funded by the European Union (ERC, UNVEIL, 101076613), and acknowledges financial contribution from PRIN-MUR 2022YP5ACE. MF is supported by a Grant-in-Aid from the Japan Society for the Promotion of Science (KAKENHI: No. JP22H01274). JDI acknowledges support from an STFC Ernest Rutherford Fellowship (ST/W004119/1) and a University Academic Fellowship from the University of Leeds. Support for AFI was provided by NASA through the NASA Hubble Fellowship grant No. HST-HF2-51532.001-A awarded by the Space Telescope Science Institute, which is operated by the Association of Universities for Research in Astronomy, Inc., for NASA, under contract NAS5-26555. CL has received funding from the European Union's Horizon 2020 research and innovation program under the Marie Sklodowska-Curie grant agreement No. 823823 (DUSTBUSTERS) and by the UK Science and Technology research Council (STFC) via the consolidated grant ST/W000997/1. CP acknowledges Australian Research Council funding  via FT170100040, DP18010423, DP220103767, and DP240103290. GL has received funding from the European Union's Horizon 2020 research and innovation program under the Marie Sklodowska-Curie grant agreement No. 823823 (DUSTBUSTERS) and from PRIN-MUR 20228JPA3A.  GR acknowledges funding from the Fondazione Cariplo, grant no. 2022-1217, and the European Research Council (ERC) under the European Union’s Horizon Europe Research \& Innovation Programme under grant agreement no. 101039651 (DiscEvol). H-WY acknowledges support from National Science and Technology Council (NSTC) in Taiwan through grant NSTC 113-2112-M-001-035- and from the Academia Sinica Career Development Award (AS-CDA-111-M03). GWF acknowledges support from the European Research Council (ERC) under the European Union Horizon 2020 research and innovation program (Grant agreement no. 815559 (MHDiscs)). GWF was granted access to the HPC resources of IDRIS under the allocation A0120402231 made by GENCI. Support for BZ was provided by The Brinson Foundation. Views and opinions expressed by ERC-funded scientists are however those of the author(s) only and do not necessarily reflect those of the European Union or the European Research Council. Neither the European Union nor the granting authority can be held responsible for them.

%\begin{acknowledgments}

%\end{acknowledgments}

\appendix

\section{Model for the rotation curve}\label{appenidx_model}
In this paragraph, we summarize the main findings of \cite{Lodato23} and \cite{martire24}. 

Although the calculations are valid for an arbitrary surface density $\Sigma$, in this work we assume that it is described by the self-similar solution of \cite{Lyndenbell74}
\begin{equation}
    \Sigma(R) = \frac{(2-\gamma)M_d}{2\pi R_c^2} \left(\frac{R}{R_c}\right)^{-\gamma}\exp\left[-\left(\frac{R}{R_c}\right)^{2-\gamma}\right],
\end{equation}
where $M_\text{d}$  and $ R_\text{c}$ are the disk mass and the scale radius respectively, $R$ is the cylindrical radius and $\gamma$ describes the steepness of the surface density, and we adopt $\gamma=1$.

The disk density at the mid-plane $\rho_\text{mid}$ is
\begin{equation}
\label{density mid}
    \rho_\text{mid} \propto \frac{\Sigma}{H_\text{mid}}\propto R^{-(\gamma+(3-q_{\rm mid})/2)}\exp\Bigg[-\Bigg(\frac{R}{R_\text{c}}\Bigg)^{2-\gamma}\Bigg],
\end{equation}
where 
\begin{equation}
    H_\text{mid}=c_{\text{s},\text{mid}}/\Omega_\text{k},
\end{equation}
is the disk hydrostatic height at the midplane, 
\begin{equation}
    c_{\text{s},_\text{mid}} =\sqrt{k_\text{b}T_\text{mid}/(\mu m_\text{p})}\propto R^{-q_\text{mid}/2}
\end{equation}
is the sound speed at the disk mid-plane, $k_{\rm b}$ is the Boltzmann constant, $\mu$ is the mean molecular weight, usually assumed to be 2.35, $m_{\rm p}$ the proton mass and 
\begin{equation}
    \Omega_{\rm k} = \sqrt{GM_\star/R^3}
\end{equation}
is the Keplerian frequency. 

We take into account that protoplanetary disks are thermally stratified by defining a function $f$ that describes how the temperature changes vertically
\begin{equation}
\begin{array}{l}
    T(R,z) = T_\text{mid}(R)f(R,z) \\
    c_s^2(R,z) = c_{s,\text{mid}}^2(R) f(R,z).
\end{array}
\end{equation}
In this work, we will use Eq. \eqref{dartois_eqt} as $f(R,z)$.
Also the density has a vertical dependence, that we describe as 
\begin{equation}
    \rho(R,z) = \rho_\text{mid}(R) g(R,z),
\end{equation}
where the value of $g(R,z)$ is linked to $f(R,z)$ through hydrostatic equilibrium. Finally, the pressure $P$ is described as 
\begin{equation}
    P(R,z)=P_\text{mid}(R)fg(R,z)=c_{\text{s},\text{mid}}^2(R)\rho_\text{mid}(R)fg(R,z).
\end{equation}
As shown in \cite{martire24}, from the hydrostatic equilibrium the relationship between $f$ and $g$ is
\begin{equation}
    \log(fg) = -\frac{1}{H_\text{mid}^2}\int_0^z \frac{z'}{f}\left[1+\left(\frac{z'}{R}\right)^2\right]^{-3/2} \text{d}z'.
\end{equation}
Hence, the density structure is
\begin{equation}
    \rho(R,z) = \frac{\rho_\text{mid}(R)}{f(R,z)}\exp\left\{-\frac{1}{H_\text{mid}^2}\int_0^z \frac{z'}{f(z',R)}\left[1+\left(\frac{z'}{R}\right)^2\right]^{-3/2} dz'\right\},
\end{equation}
that, in the isothermal case $(f=1)$ reduces to 
\begin{equation}
    \rho(R,z) = \rho_\text{mid}(R) \exp\left[-\frac{R^2}{H_\text{mid}^2} \left(1-\frac{1}{\sqrt{1+z^2/R^2}}  \right)  \right],
\end{equation}
which, for $z<<R$, reduces to the standard Gaussian profile often used to approximate the disk vertical structure. Assuming the condition of centrifugal balance, the rotation curve is given by the radial component of Navier-Stokes equation
\begin{equation}\label{vphi2}
    v_\phi^2 (R,z) =\frac{R}{\rho}\frac{dP}{dR}(R,z) + R\frac{d\Phi_\star}{dR}(R,z) + R\frac{d\Phi_d}{dR}(R,z),
\end{equation}
where $\Phi_\star$ is the stellar gravitational potential and $\Phi_d$ is the disk one. Expanding Eq. \eqref{vphi2}, we obtain 
\begin{equation}\label{rotationcurve_strat}\begin{split}
    v_\phi^2 = v_\text{k}^2 \left\{\left[1+\left(\frac{z}{R}\right)^2\right]^{-3/2} - \left[\gamma^\prime + (2-\gamma)\left(\frac{R}{R_c}\right)^{2-\gamma}  - \right. \right. \\ \left.\left. - \frac{\text{d}\log(fg)}{\text{d}\log R}\right]\left(\frac{H}{R}\right)_\text{mid}^2 f(R,z)   \right\} + v_d^2,
\end{split}\end{equation}
where  $\gamma^\prime = \gamma + (3+q_\text{mid})/2$, $v_\text{k} = \Omega_\text{k} R$ and 
\begin{equation}
	v_d^2 = G \int^\infty_0 \Bigg[K(k) - \frac{1}{4}\Bigg(\frac{k^2}{1-k^2}\Bigg)\times 
    \Bigg(\frac{R^\prime}{R}-\frac{R}{r}+\frac{z^2}{RR^\prime}\Bigg) E(k)\Bigg]\sqrt{\frac{R^\prime}{R}} k\Sigma(R^\prime) dR^\prime,
\end{equation}
where $K(k)$ and $E(k)$ are complete elliptic integrals \citep{abramowitz} and $k^2 =  4RR^\prime/[(R + R^\prime)^2 + z^2]$.

\section{DySc Code and statistical framework}\label{appenidx_dysc}
We implemented the fitting procedure for stellar mass, disk mass and scale radius based on the model of Eq. \eqref{rotationcurve_strat} in the code \textsc{DySc}\footnote{\url{https://github.com/crislong/DySc}}, already  used in \cite{Lodato23, martire24}. The code implements Markov-Chain-Monte-Carlo through the \textsc{emcee} library \citep{emcee}. According to Bayes theorem, the probability of the parameters $\theta_i = [M_\star, M_d, R_c]$, given the data $v$ with their error $\sigma_v$, and under the assumption of the model $H$ (i.e., the posterior probability $\mathcal{P}$), can be expressed as
\begin{equation}
    P(\theta_i|v;H) = \frac{P(v;H|\theta_i)P(\theta_i)}{P(v)} = \frac{P(v;H|\theta_i)P(\theta_i)}{\int \text{d}\theta_i P(v|\theta_i)P(\theta_i)},
\end{equation}
where $P(v|\theta)=\mathcal{L}$ is the likelihood, $P(\theta_i)$ denotes the priors and $P(v)=\mathcal{E}$ is the evidence. For computational reasons, it is more convenient to work with the logarithm of the probability functions. Hence, the Bayes theorem becomes
\begin{equation}
    \log\mathcal{P} = \log\mathcal{L}+\log P(\theta_i)-\log\mathcal{E}.
\end{equation}
The logarithm of the likelihood we choose is
\begin{equation}
    \log \mathcal{L} = -\frac{N}{2}\sum_i^N \log(2\pi \sigma_{v,i}) - \frac{1}{2\sigma_{v,i}^2}\left(v^\text{data}_i-v^\text{model}_i\right)^2,
\end{equation}
Here, we make the standard assumption that the data are distributed around the true value following a Gaussian distribution, with standard deviation $\sigma_v$, and that  they are not correlated. Although this is not entirely true because of the finite beam size and of the rotation curve extraction procedure, quantifying the correlation between the data is beyond the scope of the paper. The chosen priors for the model parameters are uniform distributions respectively centred on  $M_\star \in \mathcal{U} [0,5]\text{M}_\odot$, $M_d \in \mathcal{U}  [0,1]\text{M}_\odot$ and $R_c \in \mathcal{U}  [10, 1000]\text{au}$, where the lower limit for the prior is justified by the angular resolution. 

All the fits are performed fixing the power law coefficient of the surface density $\gamma=1$, underestimating the true uncertainties. In addition, this choice introduces a potential bias on the scale radius, that is the parameter most affected by the choice of $\gamma$, while the disk and stellar masses are not \citep{Andrews24}.

\section{Geometrical and thermal parameters of the sources}\label{appenidx_paramters}
Tables \ref{tab:discminer_parameters} and \ref{tab:temperature_profiles} show the emitting layer and the thermal parameters used for the fitting procedure, respectively. The heights of the emitting layers have been obtained with \textsc{discminer} \citep{Izquierdo_exoALMA} and the thermal parameters with \textsc{disksurf} \citep{Galloway_exoALMA}.

\begin{deluxetable*}{lcccccc}
\tablecaption{Height of the emitting layers extracted with \textsc{discminer} \citep{Izquierdo_exoALMA} and used in the fitting procedure in this work.\label{tab:discminer_parameters}}
\tablewidth{0pt}
\tablehead{
\colhead{\textbf{Source}} & 
\colhead{$i$ [$\deg$]} & 
\colhead{Line} & 
\colhead{$z_0$ [au]} & 
\colhead{$\psi$} & 
\colhead{$r_{\rm t}$ [au]} & 
\colhead{$q_{\rm t}$}
}
\startdata
\textbf{AA Tau} & 58.7 & $^{12}$CO $J=3-2$ & 49.8 & 1.2 & 240.1 & 1.35 \\
 &  & $^{13}$CO $J=3-2$ & 51.7 & 1.36 & 151.2 & 1.35 \\
\textbf{DM Tau} & 38.7 & $^{12}$CO $J=3-2$ & 86.6 & 1.87 & 79.6 & 0.48 \\
 &  & $^{13}$CO $J=3-2$ & 19.7 & 2.27 & 241.7 & 0.93 \\
\textbf{HD 34282} & 58.3 & $^{12}$CO $J=3-2$ & 34.0 & 1.19 & 512.2 & 3.2 \\
 &  & $^{13}$CO $J=3-2$ & 27.2 & 0.79 & 509.9 & 4.41 \\
\textbf{J1615} & 46.5 & $^{12}$CO $J=3-2$ & 26.3 & 1.04 & 529.6 & 6.89 \\
 &  & $^{13}$CO $J=3-2$ & 19.0 & 1.04 & 424.8 & 5.92 \\
\textbf{J1842} & 39.4 & $^{12}$CO $J=3-2$ & 25.9 & 1.46 & 210.6 & 1.89 \\
 &  & $^{13}$CO $J=3-2$ & 17.5 & 1.7 & 143.4 & 2.01 \\
\textbf{J1852} & 32.7 & $^{12}$CO $J=3-2$ & 75.3 & 1.78 & 60.9 & 0.84 \\
 &  & $^{13}$CO $J=3-2$ & 31.2 & 2.74 & 90.3 & 1.33 \\
\textbf{LkCa 15} & 50.3 & $^{12}$CO $J=3-2$ & 29.0 & 1.06 & 795.3 & 3.19 \\
 &  & $^{13}$CO $J=3-2$ & 27.3 & 0.87 & 511.0 & 3.46 \\
\textbf{PDS 66} & 31.9 & $^{12}$CO $J=3-2$ & 17.4 & 1.83 & 127.0 & 4.48 \\
 &  & $^{13}$CO $J=3-2$ & 7.5 & 1.2 & 29.0 & 1.54 \\
\textbf{SY Cha} & 52.4 & $^{12}$CO $J=3-2$ & 43.3 & 1.79 & 209.8 & 1.02 \\
 &  & $^{13}$CO $J=3-2$ & 72.9 & 2.44 & 66.1 & 0.7 \\
\textbf{V4046 Sgr} & 34.1 & $^{12}$CO $J=3-2$ & 25.8 & 1.84 & 151.2 & 1.17 \\
 &  & $^{13}$CO $J=3-2$ & 33.5 & 1.57 & 65.6 & 1.14 \\
\enddata
\label{table_heights}
\end{deluxetable*}

\begin{deluxetable*}{lcccccc}
\tablecaption{2-D temperature structure fits from \citet{Galloway_exoALMA} using the Dartois prescription of Eq.~\eqref{dartois_eqt}.\label{tab:temperature_profiles}}
\tablewidth{0pt}
\tablehead{
\colhead{\textbf{Source}} & 
\colhead{$T_{\rm atm, 100}$ [K]} & 
\colhead{$T_{\rm mid, 100}$ [K]} & 
\colhead{$q_{\rm atm}$} & 
\colhead{$q_{\rm mid}$} & 
\colhead{$Z_0$ [arcsec]} & 
\colhead{$\beta$}
}
\startdata
\textbf{AA Tau} & 41 & 13 & -0.51 & -0.21 & 0.45 & 0.07 \\
\textbf{DM Tau} & 37 & 20 & -0.46 & -0.37 & 0.16 & 0.0 \\
\textbf{HD 34282} & 67 & 32 & -0.0 & -0.25 & 0.28 & 0.69 \\
\textbf{J1615} & 34 & 24 & -0.1 & -0.25 & 0.21 & 1.11 \\
\textbf{J1842} & 43 & 25 & -0.45 & -0.23 & 0.18 & 0.0 \\
\textbf{J1852} & 40 & 30 & -0.87 & -0.37 & 0.11 & 0.0 \\
\textbf{LkCa 15} & 48 & 20 & -0.55 & -0.23 & 0.35 & 0.59 \\
\textbf{PDS66} & 38 & 31 & 0 & -0.08 & 0.11 & 1.21 \\
\textbf{SY Cha} & 45 & 24 & -0.58 & -0.3 & 0.31 & 0.01 \\
\textbf{V4046 Sgr} & 37 & 28 & -0.63 & -0.35 & 0.14 & 0.0 \\
\enddata\label{table_temp}
\end{deluxetable*}

\section{Best fit models and corner plots}\label{appenidx_bestfits}
\begin{figure*}
    \centering
    \includegraphics[width=\textwidth]{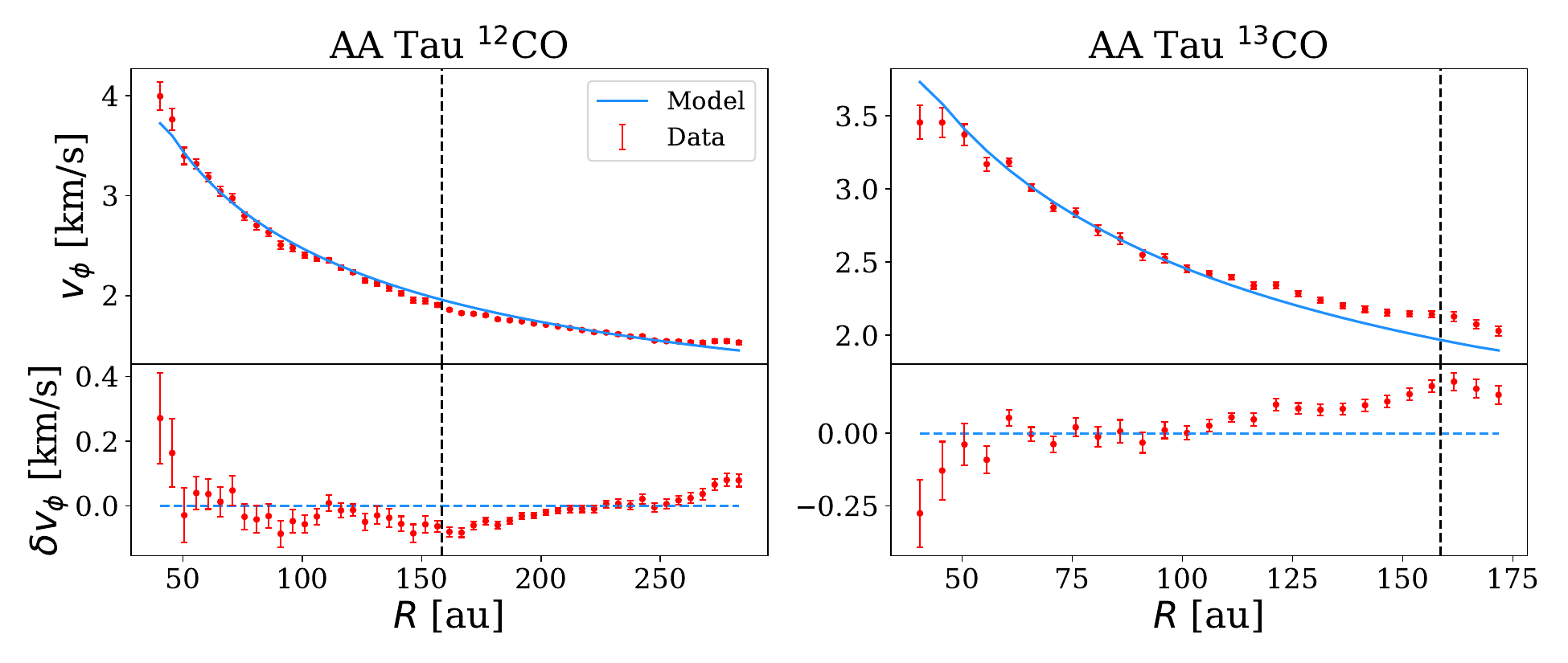}
    \includegraphics[width=\textwidth]{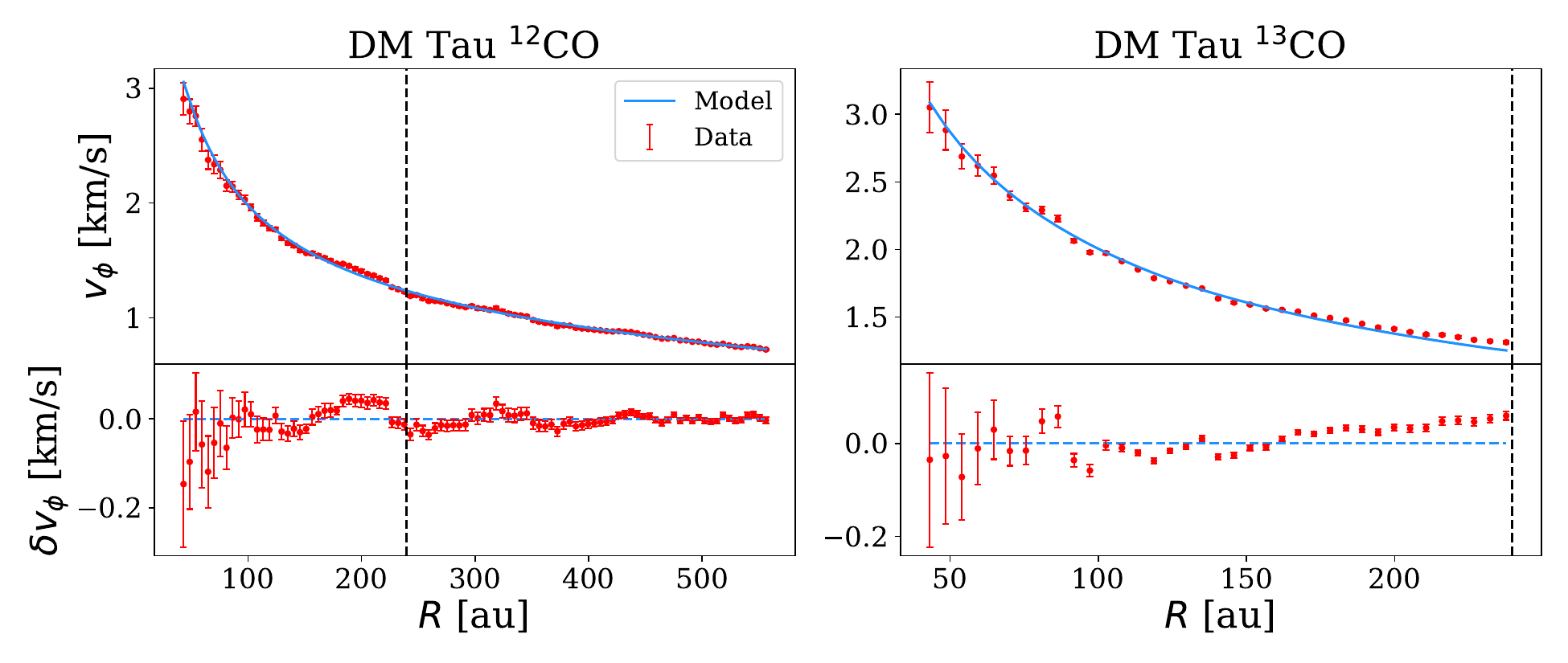}
    \includegraphics[width=\textwidth]{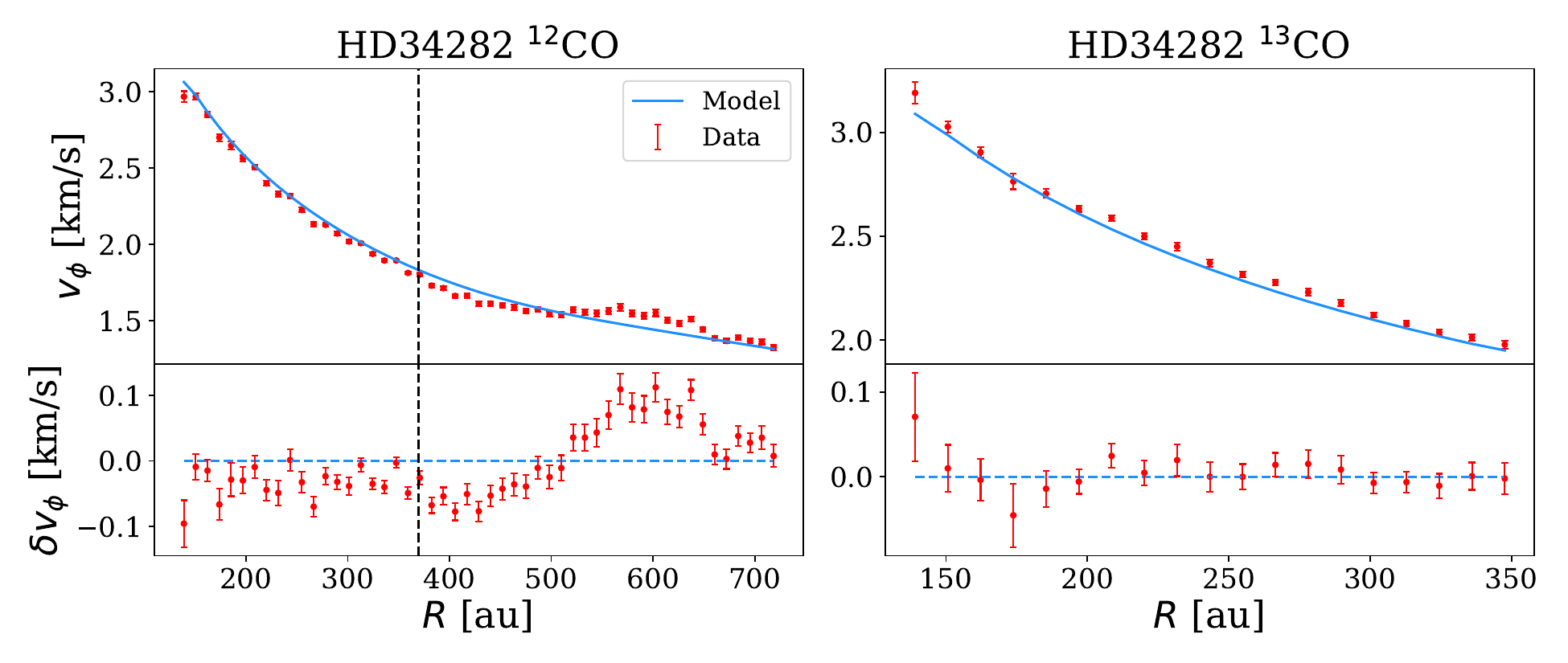}
    \caption{Rotation curve of the disks within our sample (red dots) of the $^{12}$CO (left panels) and $^{13}$CO (right panels) with the best fit model using Eq. \eqref{rotationcurve_strat}.}
    \label{rcurves1}
\end{figure*}

\begin{figure*}
    \centering
    \includegraphics[width=\textwidth]{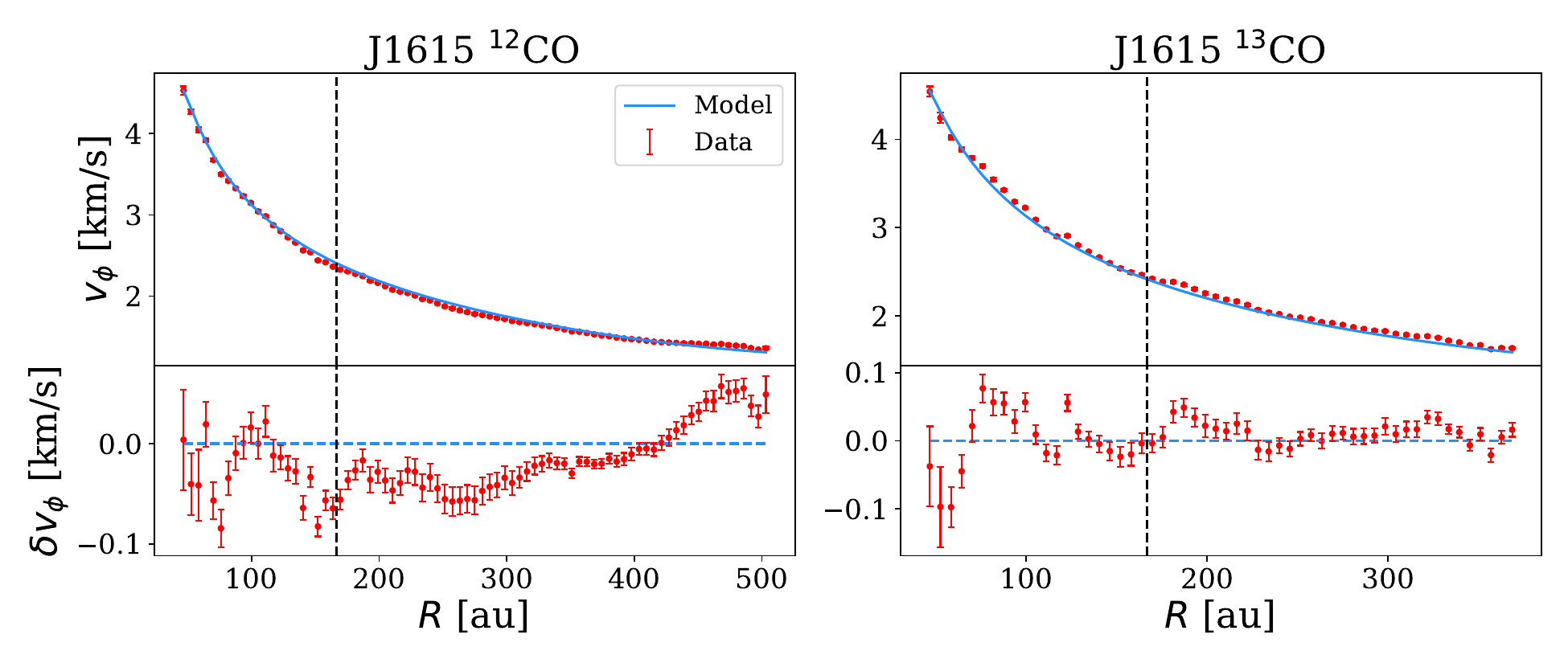}
    \includegraphics[width=\textwidth]{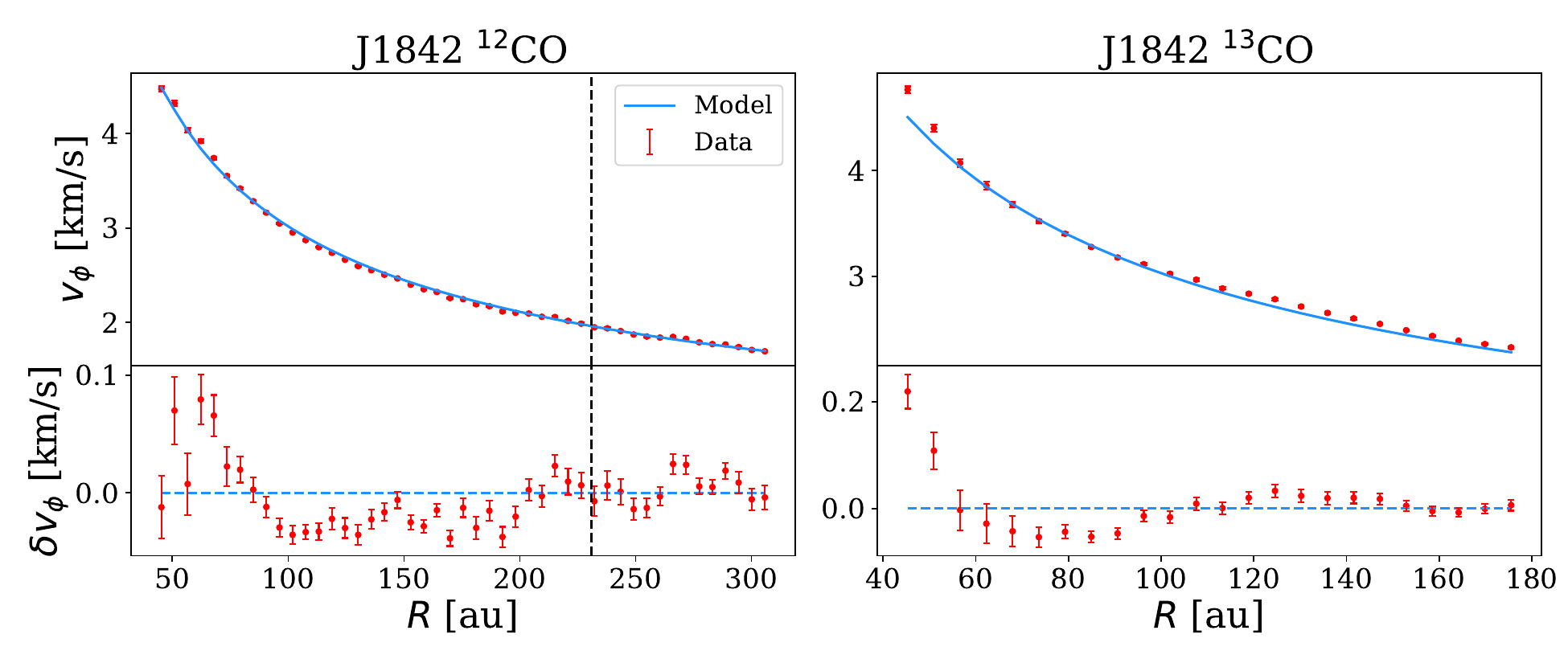}
    \includegraphics[width=\textwidth]{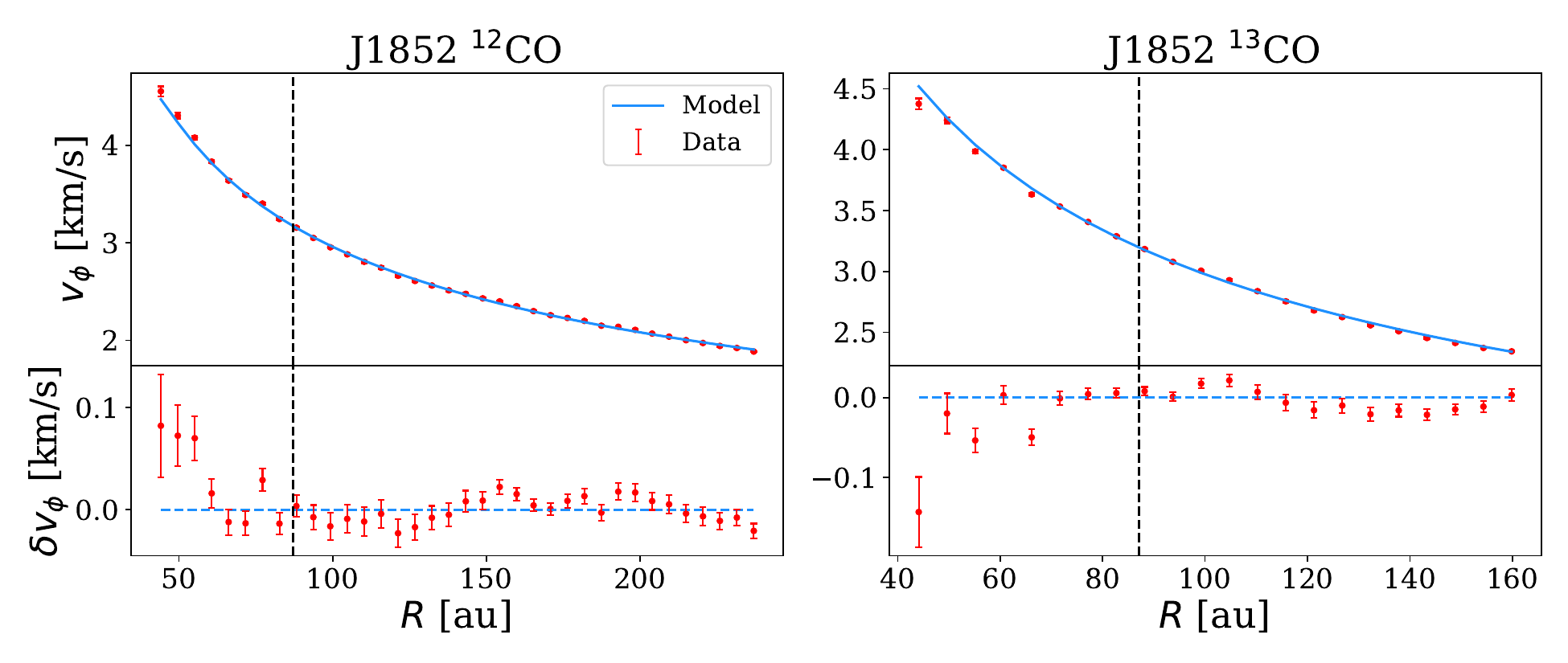}
    \caption{Rotation curve of the disks within our sample (red dots) of the $^{12}$CO (left panels) and $^{13}$CO (right panels) with the best fit model using Eq. \eqref{rotationcurve_strat}.}
    \label{rcurves2}
\end{figure*}

\begin{figure*}
    \centering
    \includegraphics[width=\textwidth]{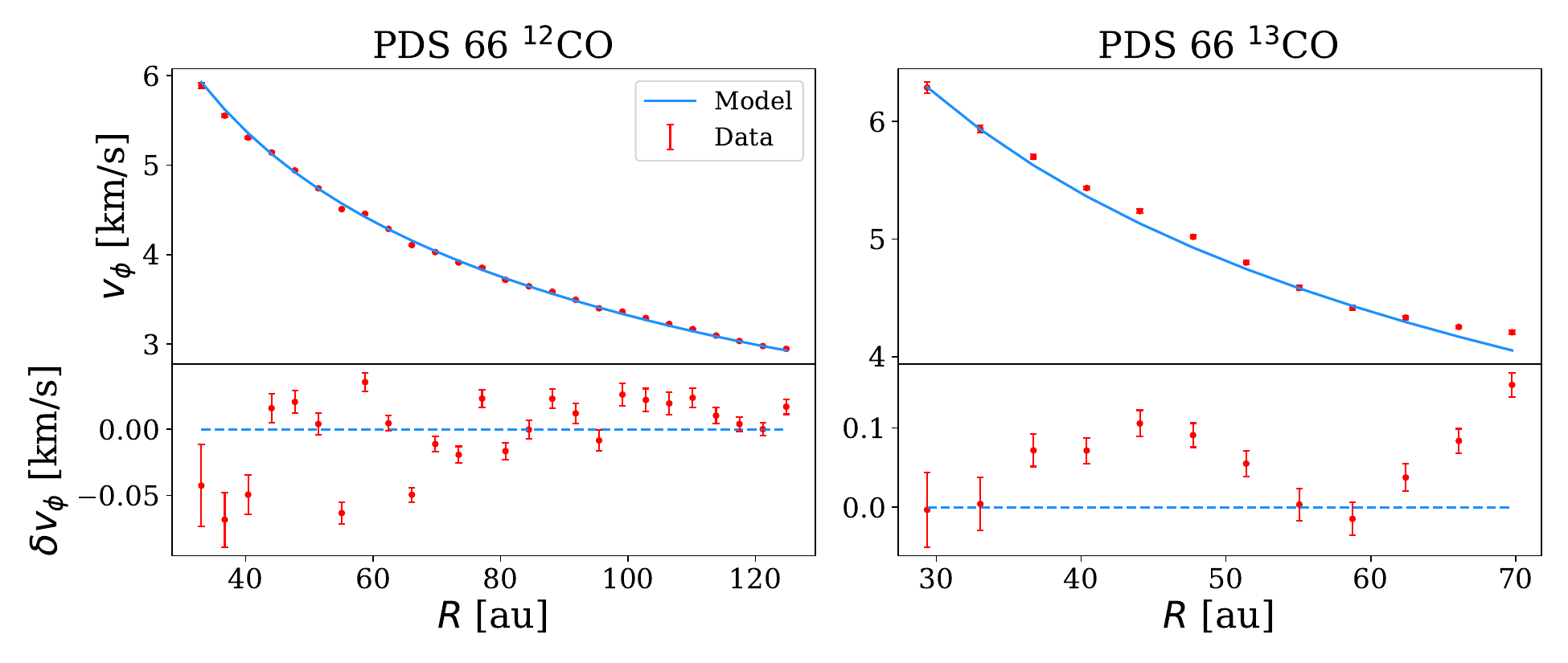}
    \includegraphics[width=\textwidth]{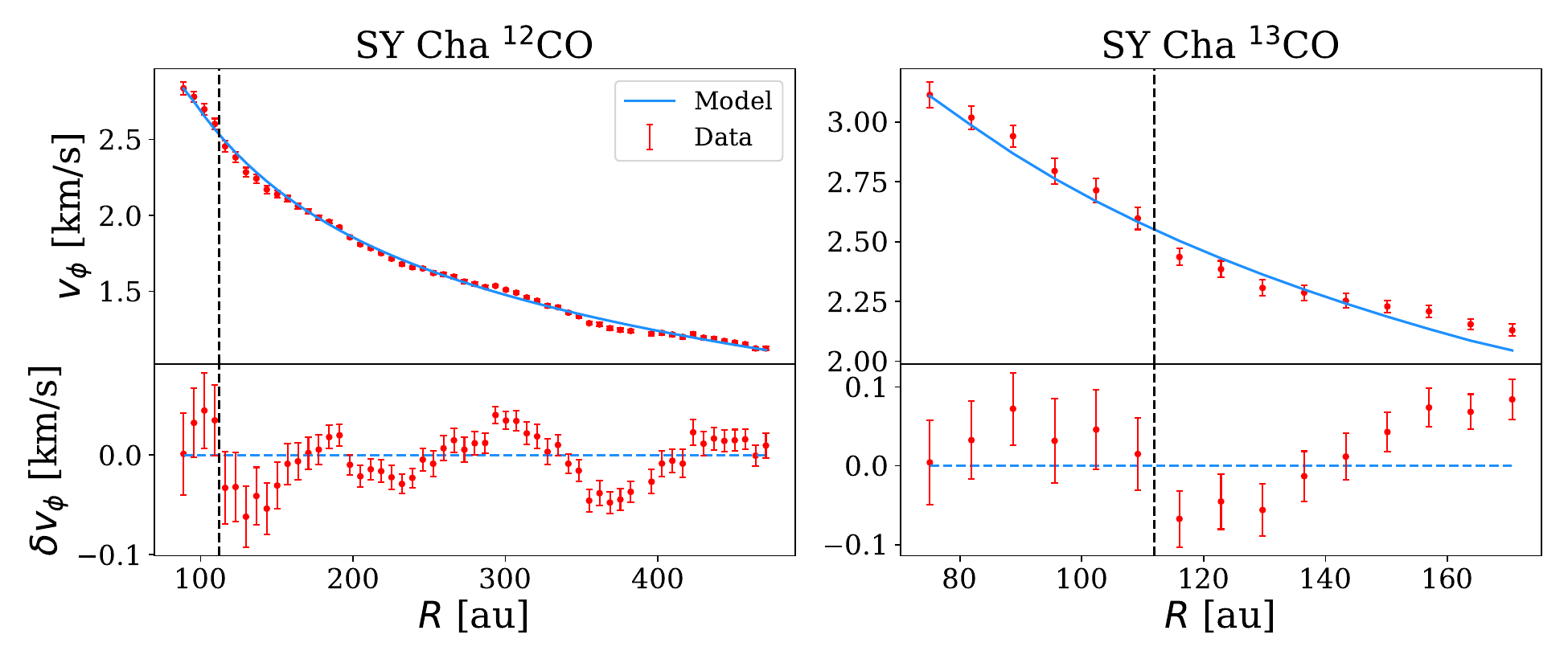}
    \includegraphics[width=\textwidth]{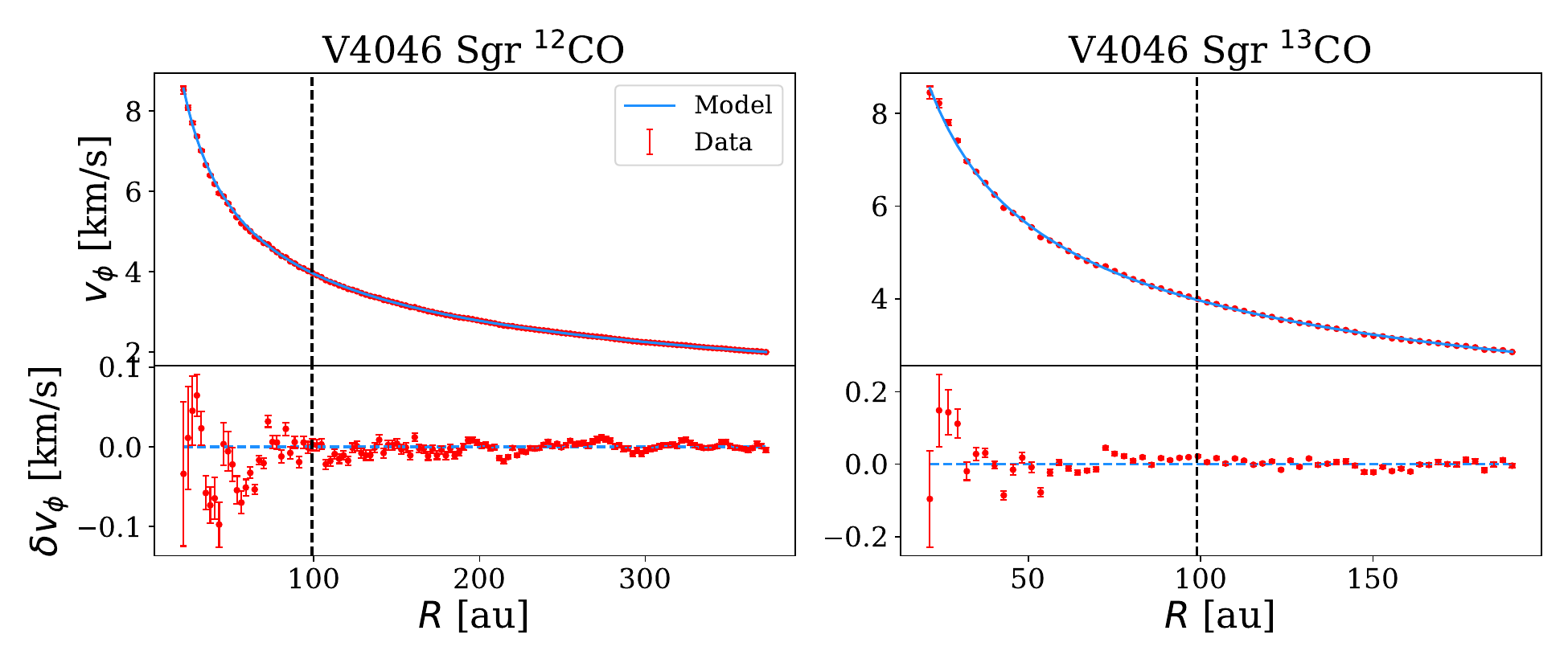}
    \caption{Rotation curve of the disks within our sample (red dots) of the $^{12}$CO (left panels) and $^{13}$CO (right panels) with the best fit model using Eq. \eqref{rotationcurve_strat}.}
    \label{rcurves3}
\end{figure*}

\begin{figure*}
    \centering
    \includegraphics[width = 0.475\textwidth]{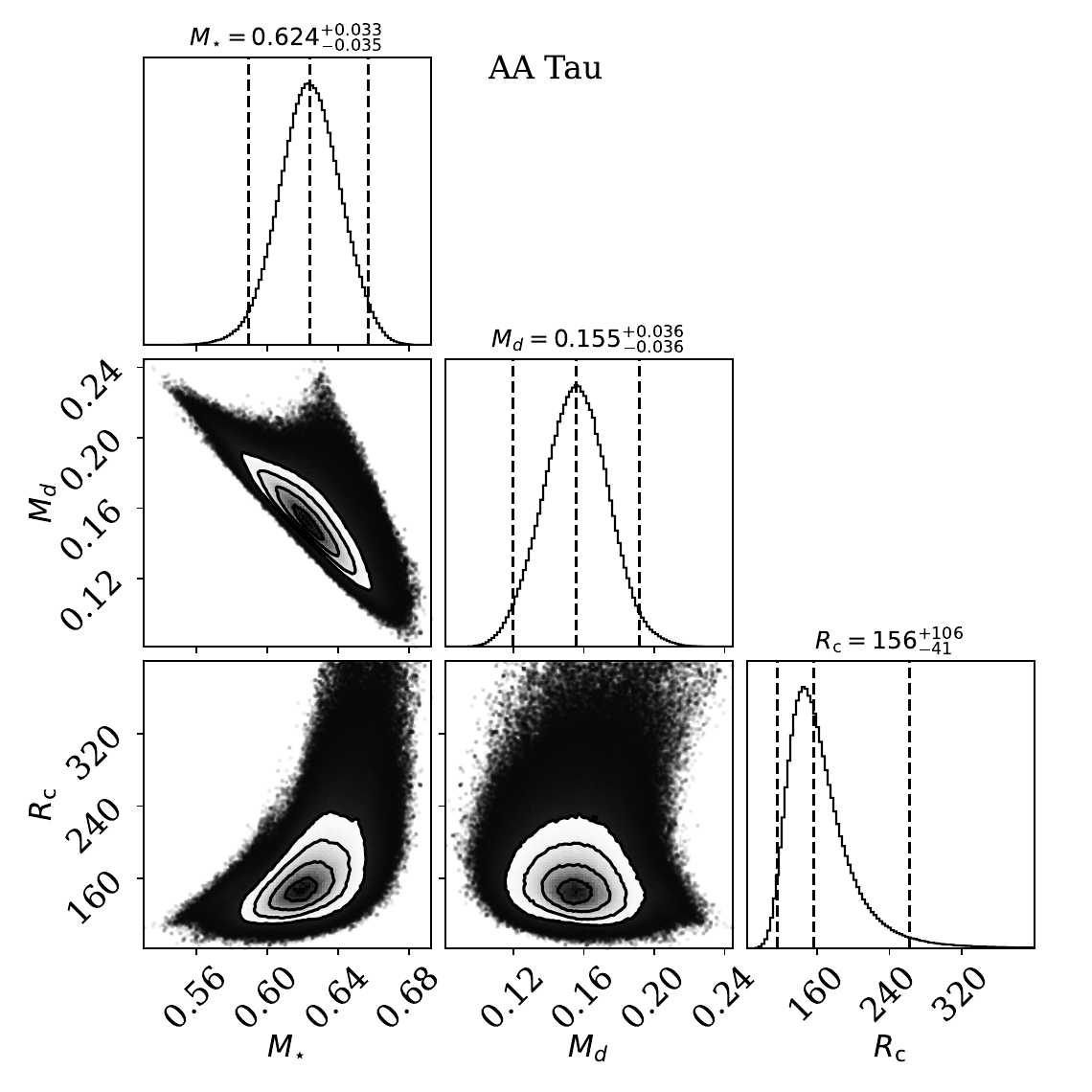}
    \includegraphics[width = 0.475\textwidth]{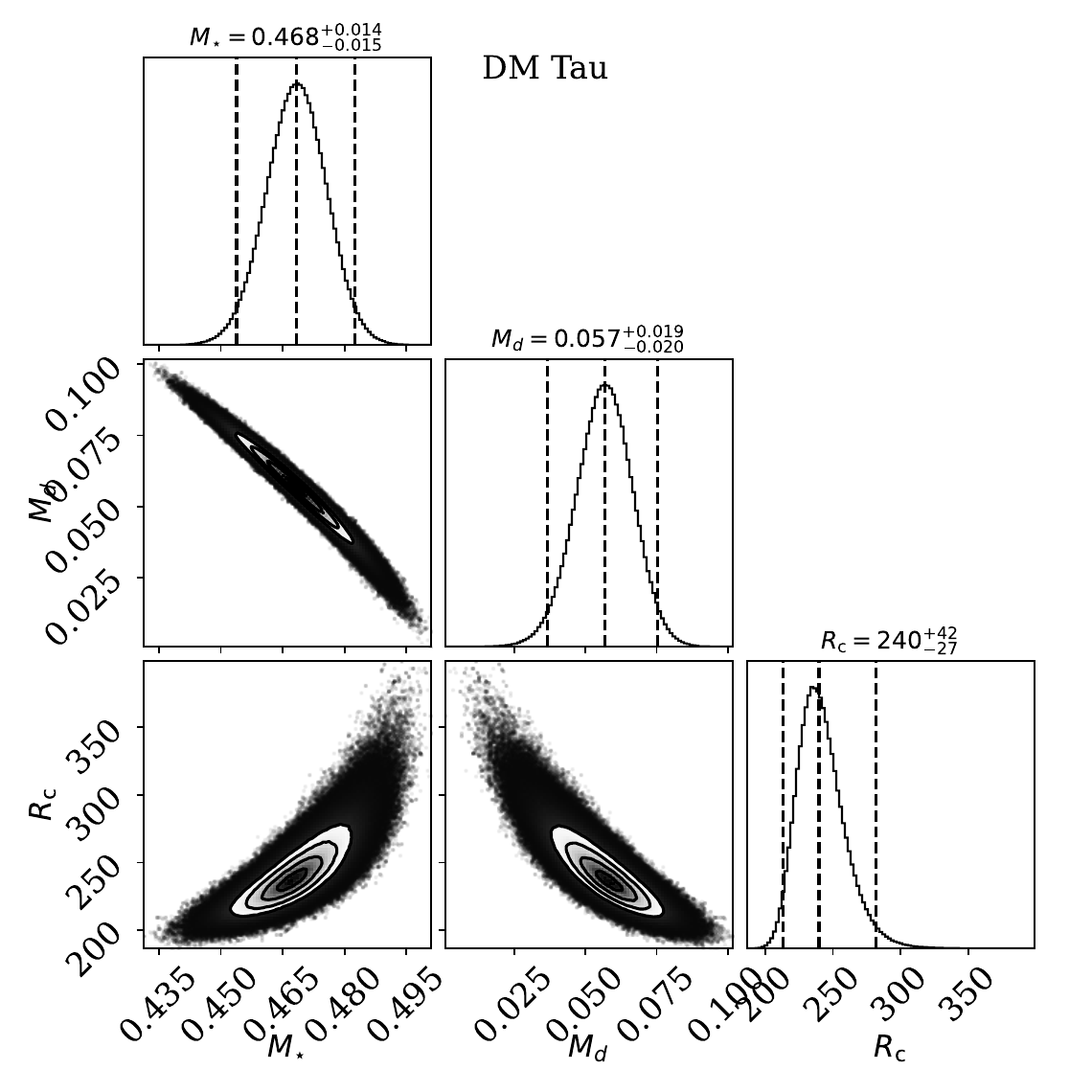}
    \includegraphics[width = 0.475\textwidth]{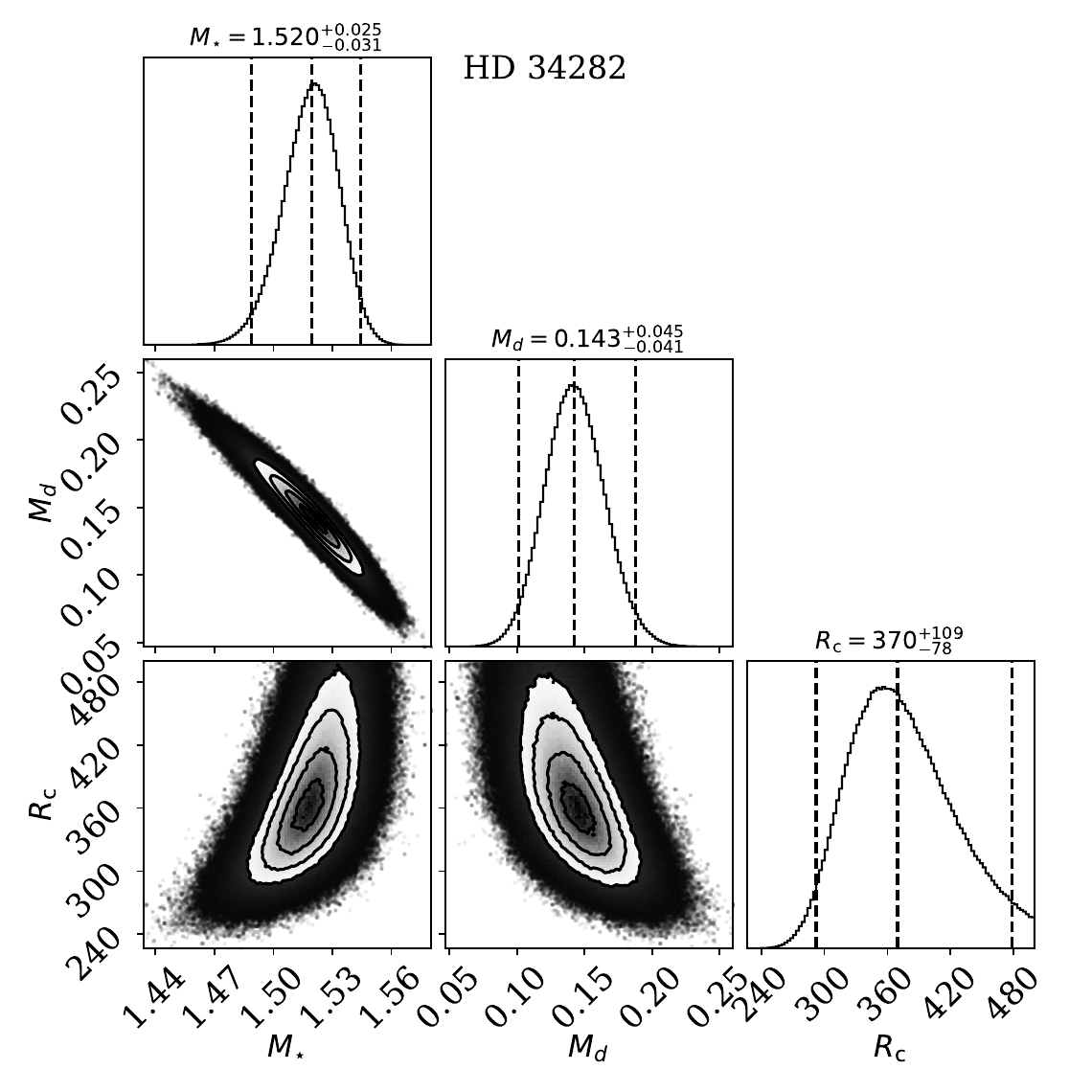}
    \includegraphics[width = 0.475\textwidth]{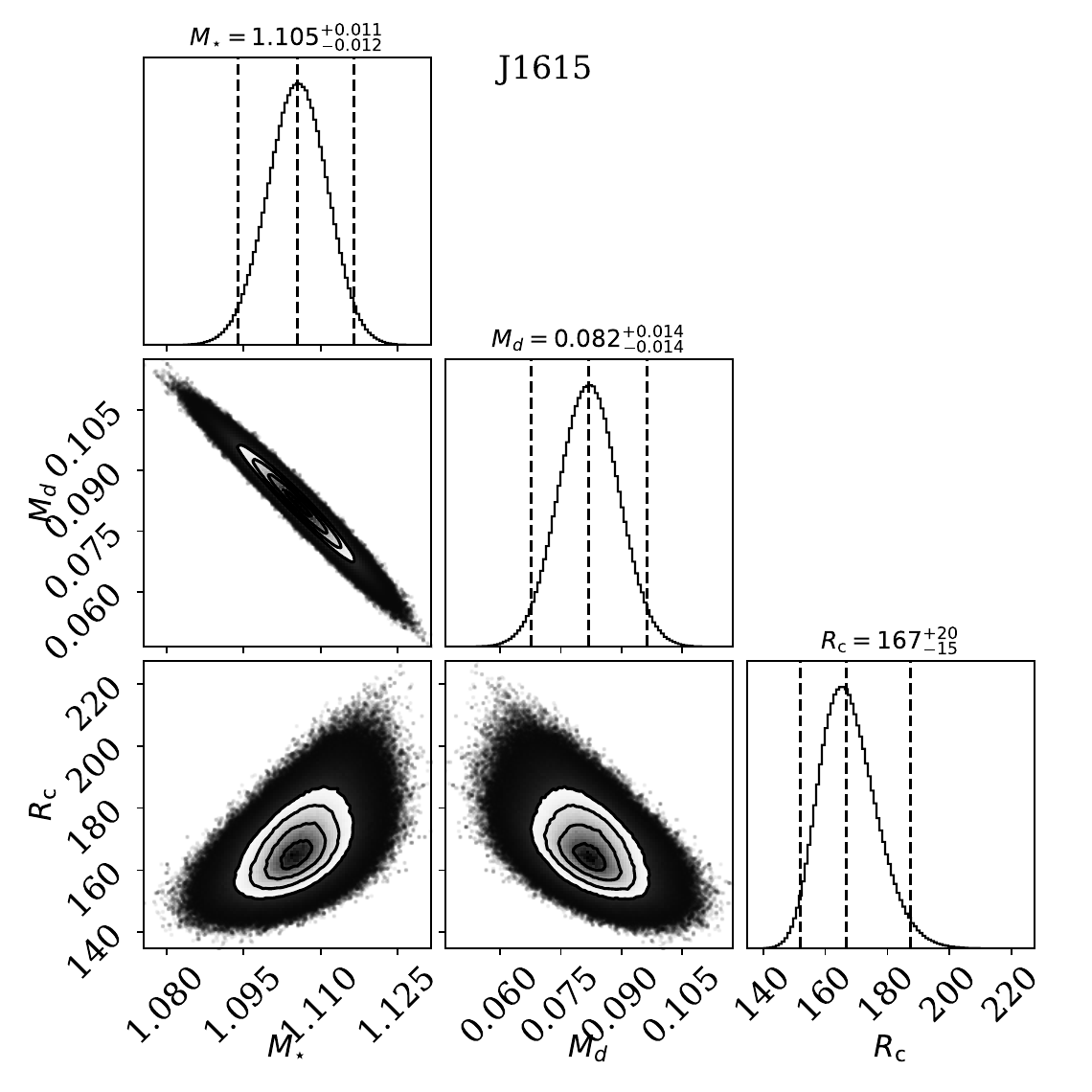}
    \caption{Corner plots for AA Tau,  DM Tau, Hd34282 and J1615.}
    \label{c1}
\end{figure*}

\begin{figure*}
    \centering
    \includegraphics[width = 0.475\textwidth]{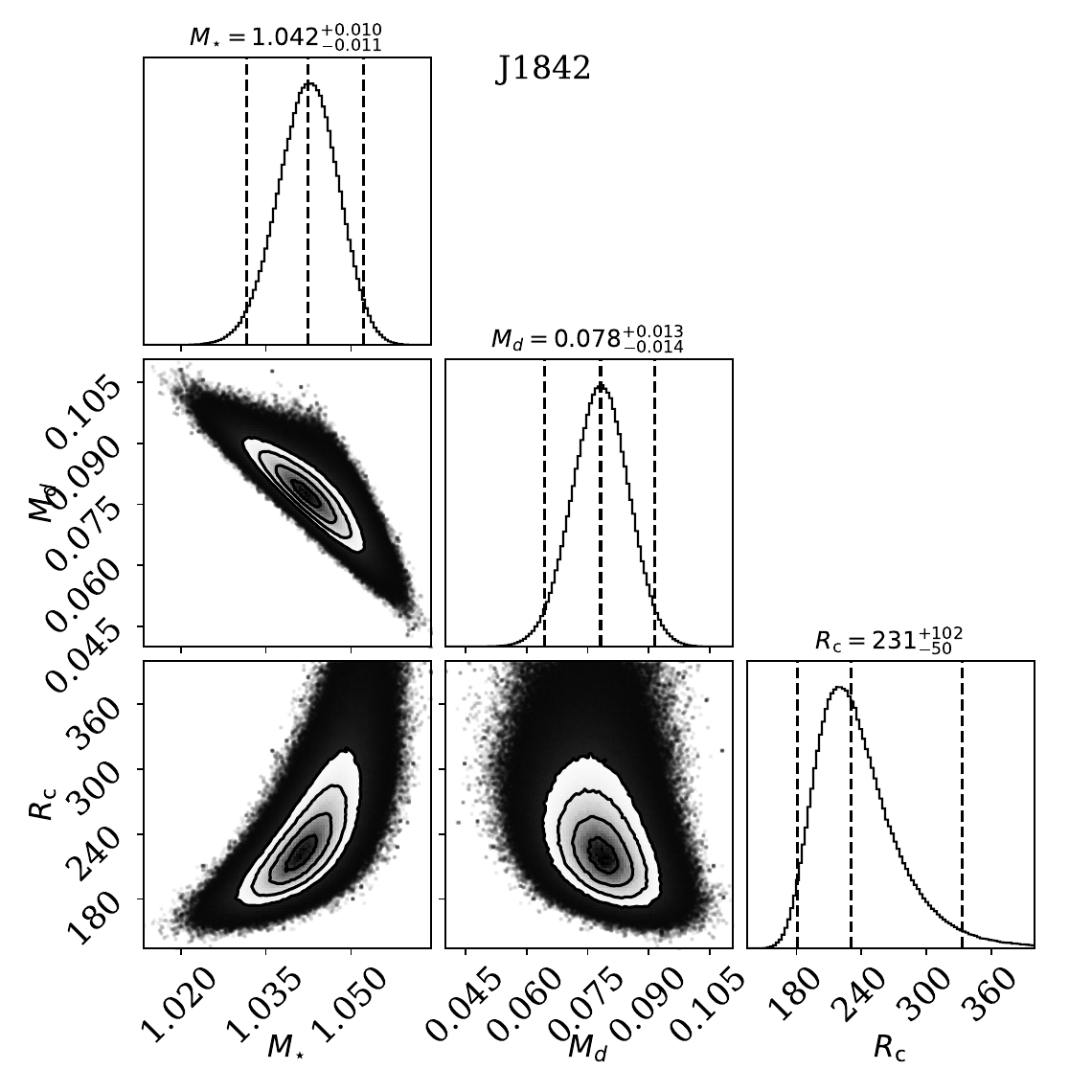}
    \includegraphics[width = 0.475\textwidth]{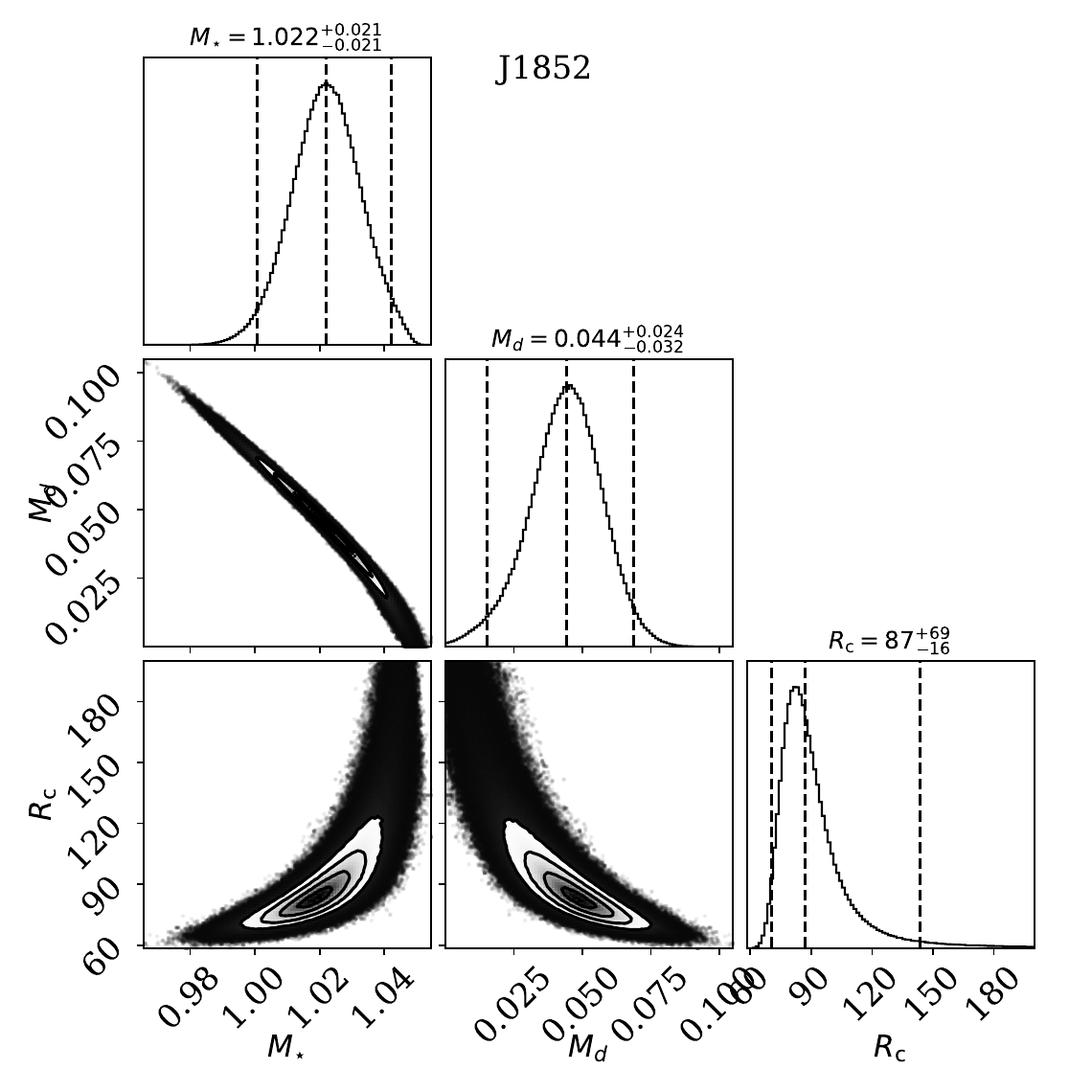}
        \includegraphics[width = 0.475\textwidth]{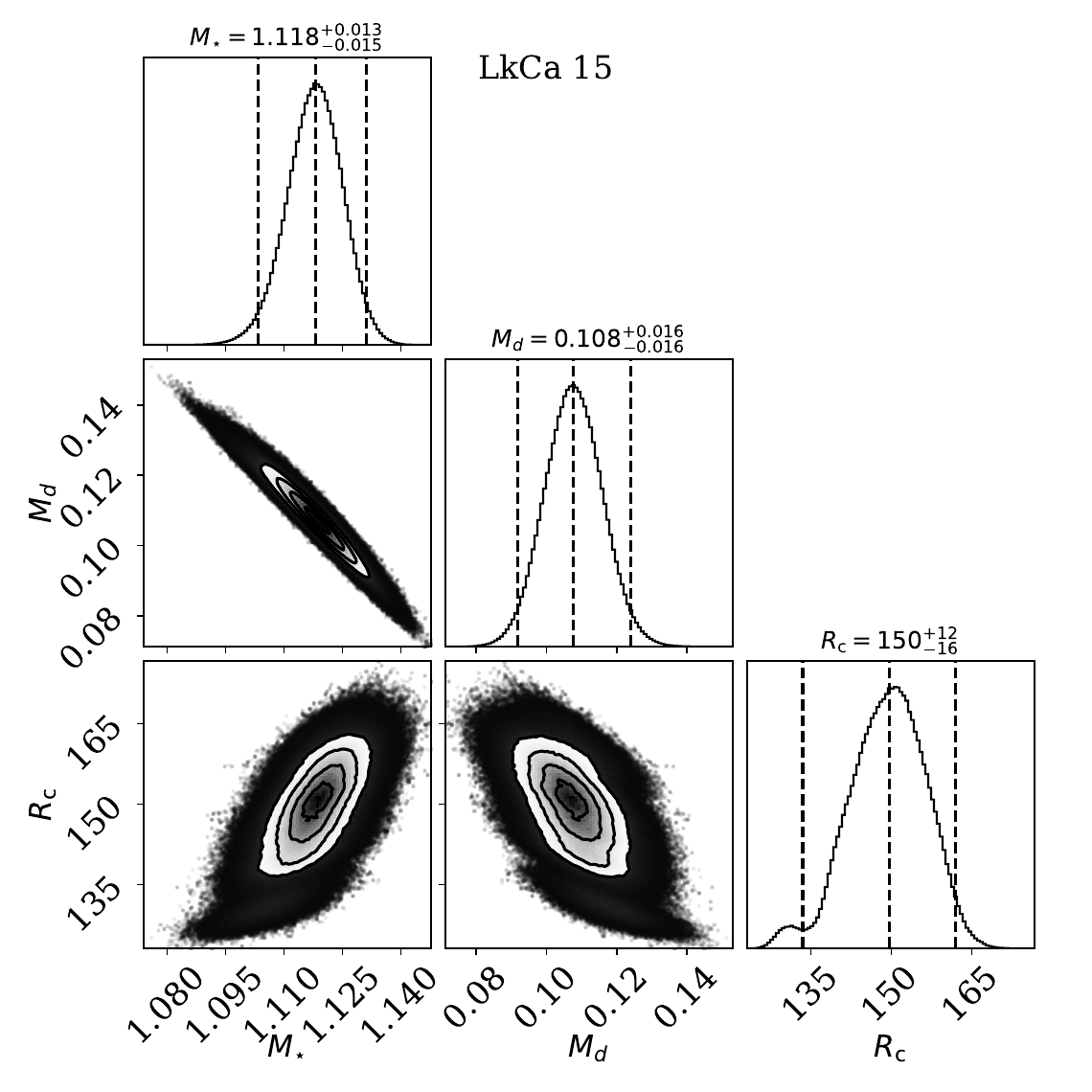}
    \includegraphics[width = 0.475\textwidth]{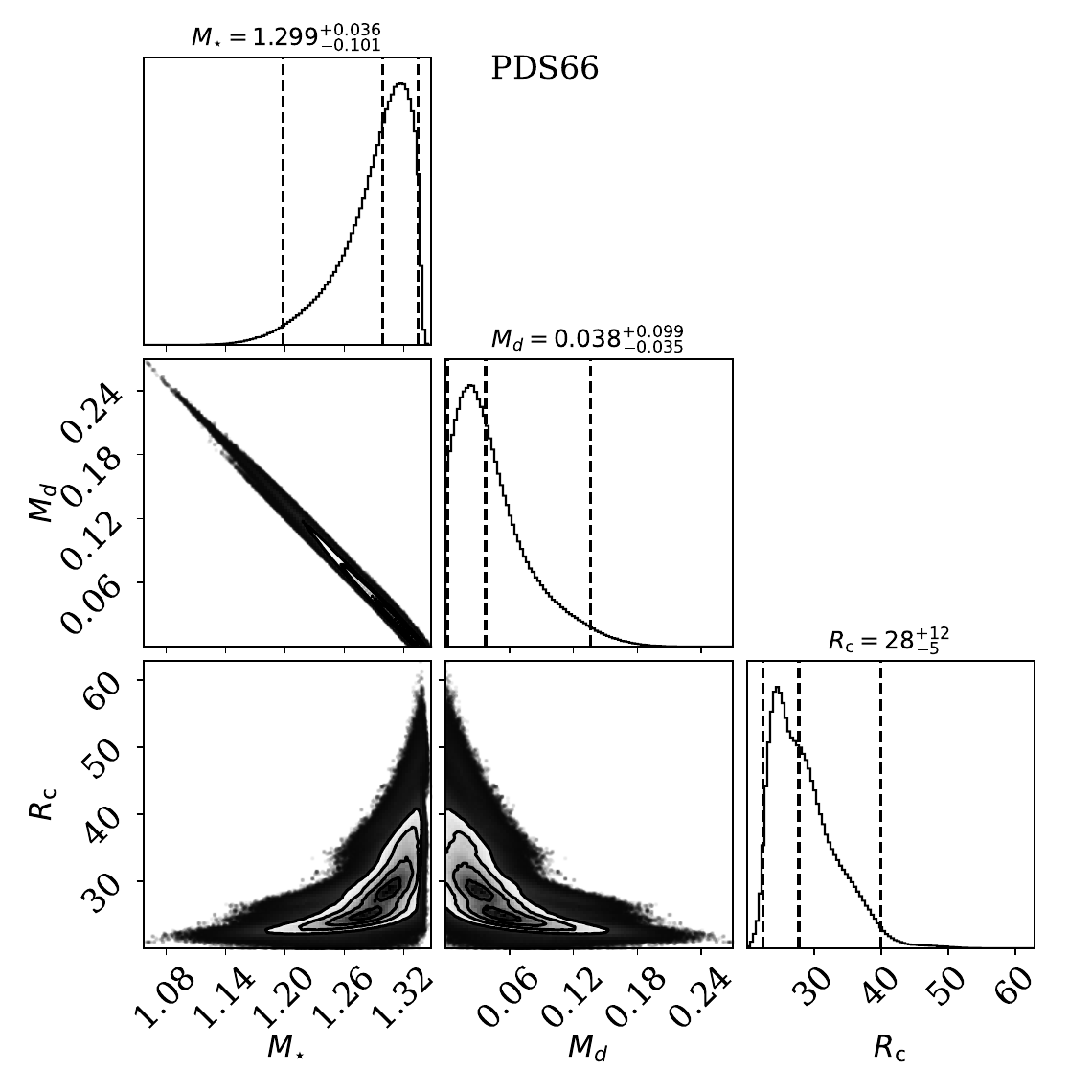}
    \caption{Corner plots for J1842, J1852, LkCa15 and PDS66.}
    \label{c3}
\end{figure*}

\begin{figure*}
    \centering
    \includegraphics[width = 0.475\textwidth]{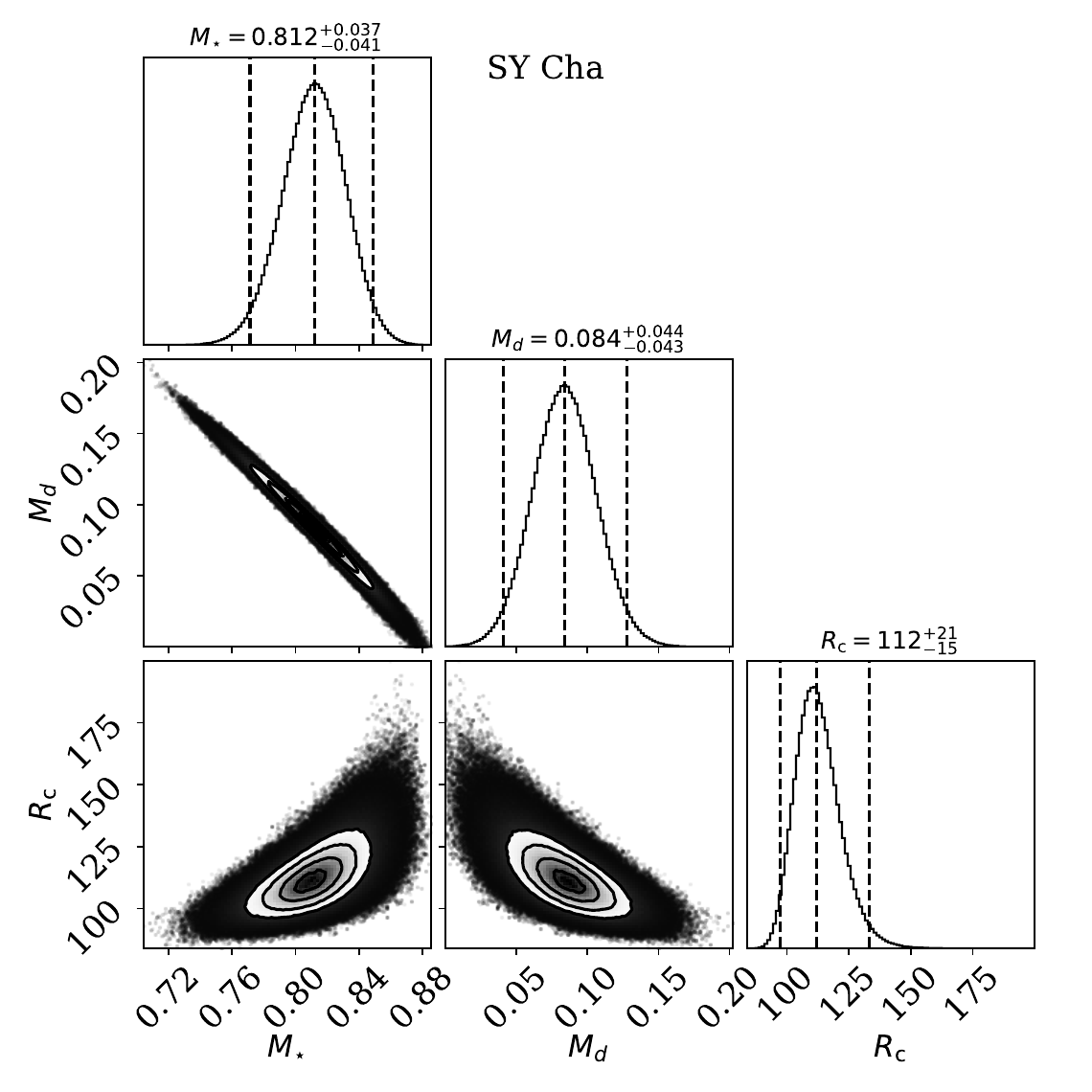}
    \includegraphics[width = 0.475\textwidth]{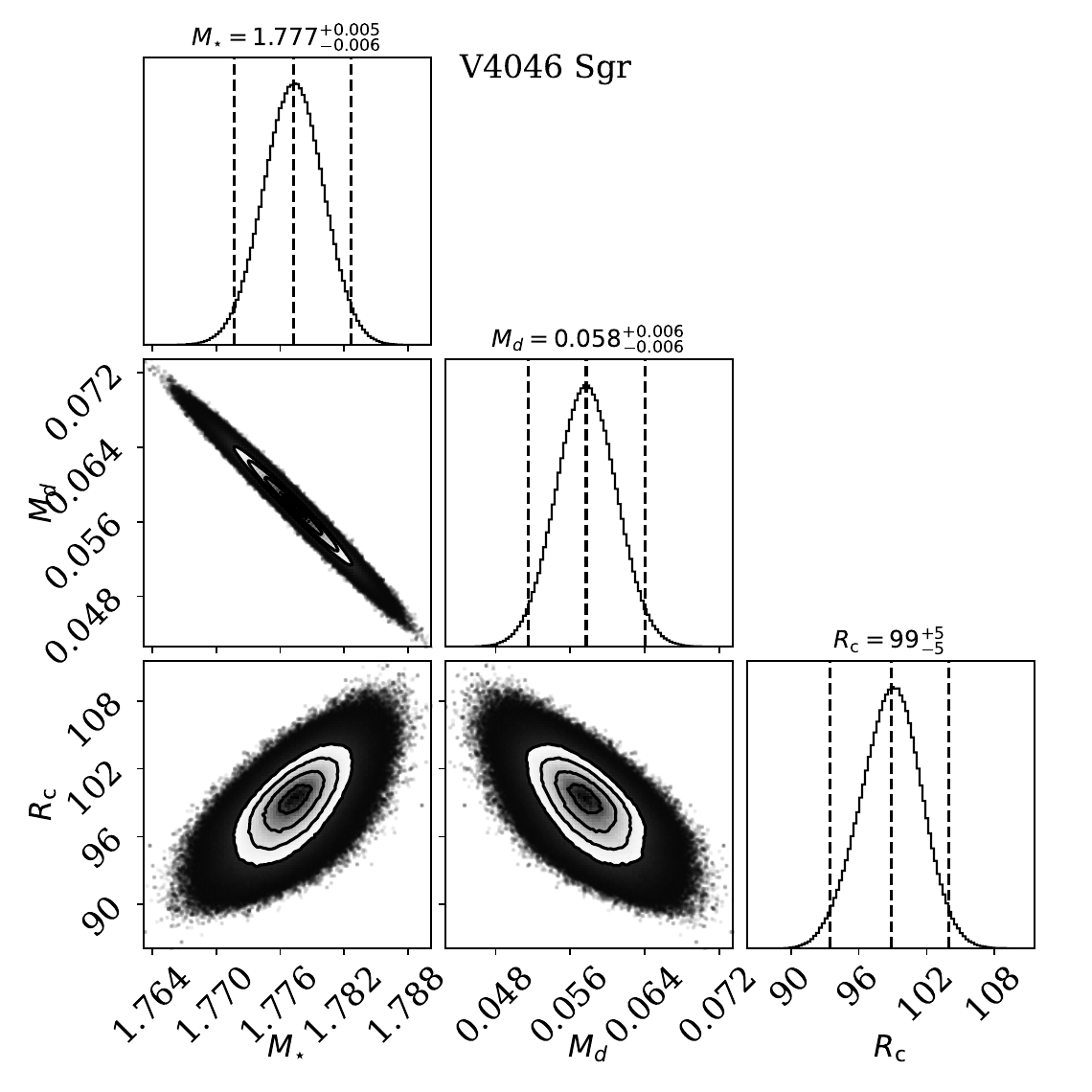}
    \caption{Corner plots for SY Cha and V4046 Sgr.}
    \label{c5}
\end{figure*}

Figures \ref{rcurves1}, \ref{rcurves2} and \ref{rcurves3} show the best fit rotation curves with the model residuals. Figures \ref{c1}, \ref{c3} and \ref{c5} show the corner plots and the posterior distributions for stellar mass, disc mass and scale radius.

\section{The case of AA Tau}\label{appenidx_aatau}
\begin{figure*}
    \centering
    \includegraphics[width=0.95\textwidth]{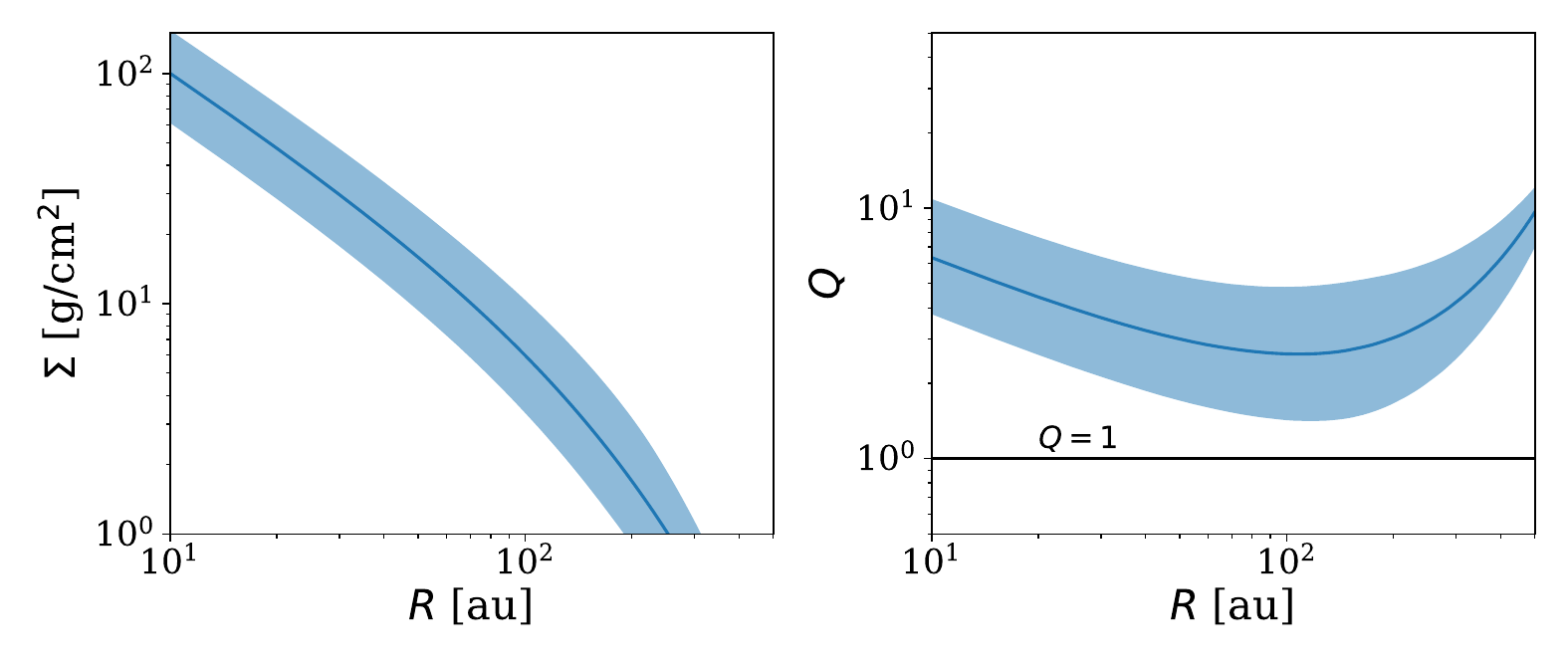}
    \includegraphics[width=0.75\textwidth]
    {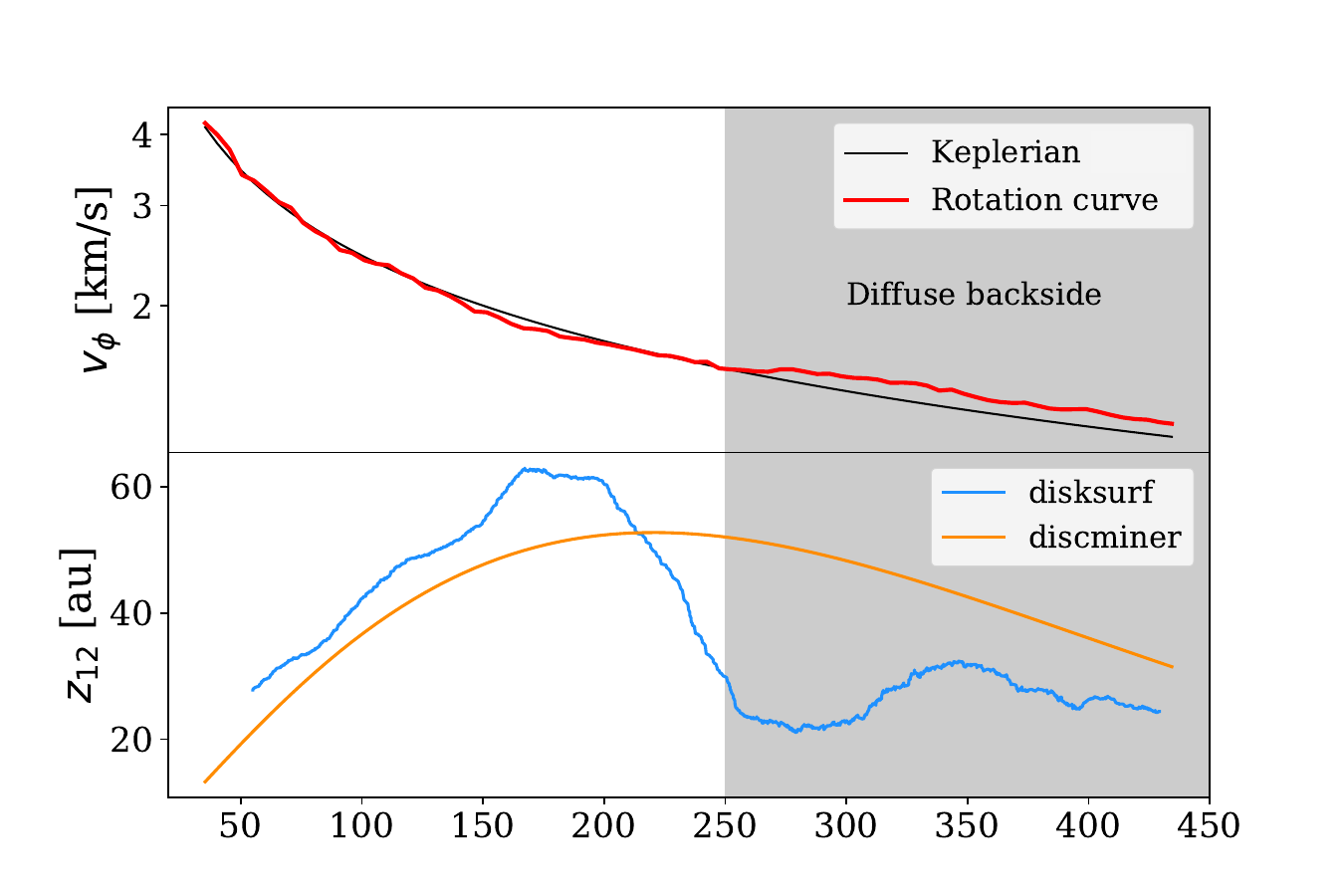}
    \caption{Top panel: Surface density and Toomre Q profiles of AA Tau, where the errorbar is computed by propagating the uncertainties on star and disk masses. Bottom panel: Rotation curve of AA Tau compared to a Keplerian curve (top panel) and comparison between the non-parametric emitting layer of \textsc{disksurf} and the parametrtic of \textsc{discminer} (bottom panel).}
    \label{aatau_problem}
\end{figure*}

As pointed out in the main text of the paper, we excluded the outer part of the $^{12}$CO rotation curve from our fits. As a matter of fact, as pointed out by \cite{Galloway_exoALMA}, the signal coming from the outer disk of AA Tau is contaminated by the backside diffuse emission. As a result, the non-parametric emitting surface extracted by \textsc{disksurf} differs from the parametric one used by \textsc{discminer}. Figure \ref{aatau_problem} shows the rotational velocity (upper panel) and the emitting height (lower panel) of the $^{12}$CO emission of AA Tau. We note that the region contaminated by diffuse backside emission $(>250\rm au)$ correspond to an increase of the rotational velocity, that becomes highly super-Keplerian. This trend is possibly due to extraction problem, and for this reason we exclude this part of the disk from our analysis.

\bibliography{bibliography}{}
\bibliographystyle{aasjournal}



\end{document}